\documentclass[11pt,a4paper]{article}
\usepackage[a4paper, margin=2.5cm]{geometry}
\usepackage{authblk}
\usepackage[utf8]{inputenc}
\usepackage[T1]{fontenc}
\usepackage{stackengine}
\usepackage{amsmath}
\usepackage{amsfonts}
\usepackage{amssymb}
\usepackage{bbm}
\usepackage{mathtools}
\usepackage{array}
\usepackage{breqn}

\usepackage{floatrow}
\usepackage{graphicx}
\usepackage{tabularx}
\usepackage{multirow}
\usepackage{wrapfig}
\usepackage{titling}
\usepackage{rotating}
\usepackage{float}
\usepackage{color}	
\usepackage[font=footnotesize,labelfont=bf]{caption}
\usepackage{subcaption}
\usepackage{hyperref}			
\usepackage{rotating}

\usepackage[numbers]{natbib}

\graphicspath{{./graphics/}}
\usepackage[normalem]{ulem}					

\def\changed#1{#1}
\def\jmochanged#1{#1}
\def\added#1{#1}
\def\addednew#1{#1}

\newcommand{\kappalin}{\kappa_{\text{\tiny lin}}}
\newcommand{\pext}{P_{\text{\tiny ext}}}

\newcommand{\Fthresh}{F^{\text{\tiny thresh}}}
\newcommand{\Finf}{F^\infty}
\newcommand{\tact}{t_{\text{\tiny act}}}

\newcolumntype{C}[1]{>{\centering\let\newline\\\arraybackslash\hspace{0pt}}m{#1}}
\newcolumntype{L}[1]{>{\raggedright\let\newline\\\arraybackslash\hspace{0pt}}m{#1}}

\title{Accounting for the geometry of the respiratory tract in viral infections}
\author{Thomas Williams$^1$, James M. McCaw$^{1,2}$ and James M. Osborne$^1$}
\date{{\footnotesize$^1$School of Mathematics and Statistics, University of Melbourne, Australia, $^2$Centre for Epidemiology and Biostatistics, Melbourne School of Population and Global Health, University of Melbourne, Australia}}

\begin{document}
	
	\maketitle

	\begin{abstract}
		Increasingly, \changed{experimentalists and modellers alike} have come to recognise the important role of spatial structure in infection dynamics. Almost invariably, spatial \changed{computational} models of viral infections \changed{--- as with \emph{in vitro} experimental systems --- represent the tissue as wide and flat,} which is \added{often} assumed to be representative of \changed{entire affected tissue within the host. However,} this assumption fails to take into account the distinctive geometry of the \added{respiratory tract} 
		\changed{in the context of 
			viral infections}. The \added{respiratory tract} 
		is characterised by a tubular, branching structure, and moreover is spatially heterogeneous: deeper regions of the lung are composed of far narrower airways and are associated with more severe infection. Here, we extend a typical multicellular model of viral dynamics to account for two essential features of the geometry of the \added{respiratory tract}: 
		the tubular structure of airways, and the branching process between airway generations. We show that, with this more realistic tissue geometry, the dynamics of infection are substantially changed compared to  \jmochanged{standard computational and experimental approaches}, and that the resulting model is equipped to tackle important biological phenomena that \changed{do not arise in a flat host tissue}, including viral lineage dynamics, 
		and heterogeneity in immune responses to infection in different regions of the respiratory tree. \changed{Our findings suggest aspects of viral dynamics which current \emph{in vitro} systems may be insufficient to describe, and points to several features of respiratory infections which can be experimentally assessed.}
		\vspace{1em}
		
	\end{abstract}

	\section{Introduction}
	
	Respiratory viruses represent a widespread and ongoing threat to global public health, both through seasonal epidemics of viruses like influenza or respiratory syncytial virus (RSV), and pandemic viruses like SARS--CoV--2 and newly emergent influenza strains. Within the host, viral infections of the respiratory tract can have highly varied patterns of progression, and can lead to marked differences in pathogenicity. In general, milder illnesses tend to arise from infections in the upper airways, such as the nasopharynx and trachea, whereas more severe and potentially life--threatening illnesses typically involve infection of the lower airways and alveolar spaces of the deep lung \citep{ke_perelson_et_al_covid_upper_and_lower_rt, wolfel_et_al_virological_assessment_covid, gallagher_spatial_spread, taubenberger_and_morens_lung_influenza_pathology}. Mathematical models of viral infections within the host have offered important insights into the dynamics of respiratory viruses, including how these infections spread in space \citep{smith_and_perelson_influenza_review, gallagher_spatial_spread, bauer_et_al_agent_based_models_virus}. However, traditional approaches to modelling the spatial dynamics of viral infections --- including infections of the \added{respiratory tract} 
	--- do not consider the specific geometry of the tissue. Instead, models typically assume a flat, roughly square tissue with either periodic or no-flux boundary conditions \citep{sego_et_al_covid_model, levin_et_al_T_cell_search_influenza, whitman_et_al_spatiotemporal_host_virus_influeza}. \changed{This approach mirrors that of \emph{in vitro} infection assays. In both cases, these model systems (computational and experimental) \added{often} rely on the assumption that the chosen model domain represents a small, typical patch of the overall tissue within the host}. Implicit in this setup is the assumption that the overall tissue is reasonably wide and homogeneous, at least compared to the size of the patch. However, this is not necessarily true of the \added{respiratory tract}. 
	The lung is an extremely structurally complex organ, comprising 24 generations of branching airways ranging in circumference from centimetres in diameter to fractions of millimetres in the alveolar ducts \citep{chen_sars_cov_2_in_lung, makevnina_lung_atlas}. The epithelial lining of the \added{overall} respiratory tree --- targeted by respiratory viruses --- therefore occupies an extremely large, intricately--structured surface area within a highly compact volume \citep{chen_sars_cov_2_in_lung, taubenberger_and_morens_lung_influenza_pathology}. As such, while a flat, square model tissue may offer a reasonable approximation of the structure of the upper airways, it fails to represent the extremely narrow, highly branched structure of the lower airways, which are primarily targeted by more highly pathogenic viruses \citep{gallagher_spatial_spread, ke_perelson_et_al_covid_upper_and_lower_rt}. 
	
	Two recent \changed{computational modelling} studies have attempted to address this limitation by using models of infection spread in space which explicitly represent structural features of the \added{respiratory tract}. 
	Chen and colleagues studied viral infection dynamics on individual airways from different generations of the lung \citep{chen_sars_cov_2_in_lung}. Their model explicitly included the mucus layer lining the susceptible tissue as the principal medium in which virions spread, and incorporated experimental measurements of its velocity at different depths in the lung. Their modelling work suggested that, based on the properties of the mucus, there was a pronounced heterogeneity in the likelihood of an infection being established in different generations of the lung. Another study by Moses and coworkers compared the dynamics of SARS--CoV--2 infection on a large two--dimensional sheet of cell to the dynamics on a full, three-dimensional computational model of the entire human \added{respiratory tract} 
	\citep{moses_et_al_lung_covid_model}. The authors found that infection on the \added{realistic, three--dimensional} 
	geometry was notably accelerated compared to infection of the two--dimensional tissue, and resulted in an enhanced virion production. These two works hint at the heterogeneity between the dynamics of different regions of the \added{respiratory tract}, 
	and suggest that there are important distinctions between the dynamics of infection on tissues with realistic 
	geometry, and more standard, flat model tissues. However, due to the complexity of \added{the models used by both sets of authors,} it is not clear how the tissue geometry, \added{specifically}, influences the dynamics of the infection, or what predictive capacity \added{may be conferred to models which account for this in a more realistic way}. There is therefore a need for a more \added{focused} analysis, \added{using models of reduced complexity,} of the role of the structure of the \added{respiratory tract} 
	in spatial models of respiratory viral infections. 
	
	In this work, we seek to explore \changed{through simulation study} how the two main features of \added{respiratory tract} 
	geometry --- the tubular and branching tissue structure --- influence the dynamics of a multicellular model of viral infection. We simulate our model on computational tissues with varying 
	geometries \added{representative of the respiratory tract} to explore the role that different structural properties play in influencing infection dynamics. Our results 
	\added{show that the inclusion of realistic features of tissue geometry is sufficient to generate infection dynamics which are substantially different than those on flat, unstructured tissues}. \added{This suggests that the structure of the respiratory tract may have a substantial impact on the way respiratory viral infections progress within the host, and may have broader relevance to infections by other types of pathogens such as bacteria.} \added{Our findings} demonstrate a \jmochanged{clear} need for improved experimental systems capable of representing 
	\added{the dynamics of respiratory infections \emph{in situ}}. Aided by the future generation of appropriate data, we show that 
	\added{models like ours --- combined with the incorporation of more detailed descriptions of pathogen--host interactions --- will be necessary} to address important biological questions on the dynamics of respiratory infections \emph{in vivo}.

	\section{Methods}\label{sec:methods}
	
\begin{figure*}
		\centering
		\includegraphics[width=0.8\textwidth, page=1, trim={2.5cm, 10cm, 2.5cm, 10cm}, clip]{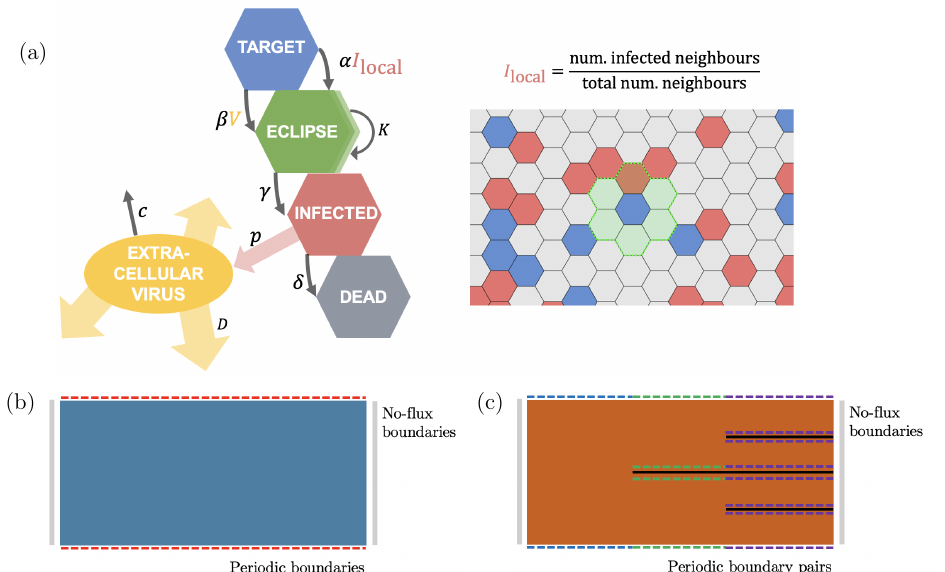}
		\caption{(a) Schematic illustrating the basic components and dynamics of the model used in this work. (b), (c) Sketch of the tube and branching geometry, respectively, as used in this work. \added{Pairs of dashed lines of the same colour indicate pairs of periodic boundaries, while solid black lines in (c) represents cuts in the sheet.}}
		\label{fig:model_cartoon}
	\end{figure*}

	In this work, we extend a relatively simple and established agent-based model based on one which we and others have used in earlier works \citep{williams_et_al_inference, blahut_et_al_hepatitis_c_two_modes_of_spread, durso_cain_hcv_dual_spread}, \added{and which is fully described in Supplementary Section S1}. This model couples a grid of discrete, spatially--explicit \added{epithelial} cells to a viral density surface. We assume that, at a given time, cells are in one of the following states: susceptible to infection (\emph{target}), infected but not yet infectious (\emph{eclipse}), productively infectious (\emph{infected}), or dead (\emph{dead}). Infection can arise from either cell--free virions at rate $\beta$ --- secreted by productively infectious cells at rate $p$ --- or from direct cell--to--cell infection between susceptible and productively infectious cells at rate $\alpha$. Latently infected cells undergo an eclipse phase of gamma--distributed duration with mean $1/\gamma$ and shape parameter $K$. In line with experimental observations for SARS--CoV--2, we assume that the cell-to--cell infection mechanism is the dominant mode of infection and tune our model parameters such that approximately 90\% of infections arise from this mechanism; the other parameters are taken from fitting an 
	\added{analogous} ODE model to experimental data \added{for influenza infection \emph{in vitro}}. For full details on parameter selection, refer to Supplementary Section S2. Extracellular viral density diffuses across the tissue according to linear diffusion with coefficient $D$ and uniformly decays in the environment at rate $c$. \added{This model and its parameterisation were developed with SARS--CoV--2 and influenza in mind, but are readily generalisable to other respiratory viruses.} We show a schematic of this model setup in Figure \ref{fig:model_cartoon}(a). For full details of the model and its implementation, refer to the Supplementary Information.
	
	We simulate our model in either tubular or branching geometries. In each case, the unrolled tissue is a 2D, hexagonally--packed grid of cells (consistent with the packing of real epithelial monolayers \citep{holder_et_al_design_considerations_influenza,beauchemin_et_al_modeling_influenza_in_tissue,blahut_et_al_hepatitis_c_two_modes_of_spread}), where the geometry is implemented in the boundary conditions of the sheet. For the tube geometry, we impose periodic boundary conditions on the top and bottom edges with no-flux boundary conditions on the left and right edges of the tissue, which we show in Figure \ref{fig:model_cartoon}(b). \added{No-flux (as opposed to periodic) boundary conditions are imposed on the ends of the tube to ensure that --- in keeping with the biological reality --- infection cannot jump from one end of the tube to the other. This choice of boundary conditions, moreover, prevents any viral density from leaving the system.} 
	
	Tissues with branching geometry are essentially an ensemble of tubes of varying dimensions. We assume that the tissue undergoes even, binary division at a sequence of branching points along the horizontal. Each offspring tube has periodic boundary conditions on its top and bottom edge and is assumed to be half the width (circumference) of its parent, which we show in Figure \ref{fig:model_cartoon}(c). This ensures that the overall shape of the unrolled branching tree maintains constant width. We impose zero--flux boundaries on the \added{ends of the} overall branching tree. For full details of the implementation of the tissue geometries, refer to Supplementary Sections S3--S5.

	\section{Results}
	
	\subsection{\addednew{Dependence of viral dynamics on the rate of viral diffusion is increased on tubular tissues}}\label{sec:diffusion_matters}

\begin{figure*}[h!]
		\centering
		
		\includegraphics[width=0.82\textwidth, page=2, trim={2.5cm, 11.5cm, 2.5cm, 2.5cm}, clip]{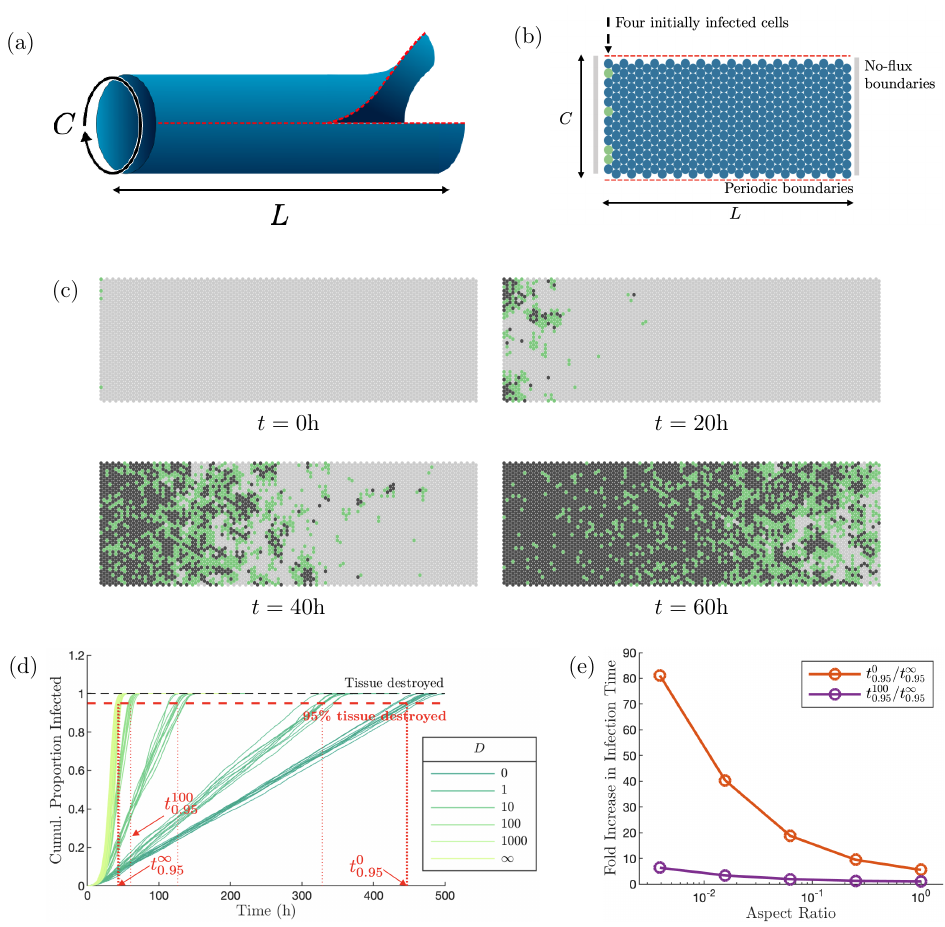}
		
		\caption{(a) Schematic illustrating our implementation of a tube-shaped tissue, as the surface of a cylinder of cells of circumference $C$ and length $L$. (b) The unrolled tissue depicted in (a), indicating the periodic horizontal boundaries and no-flux vertical boundaries, along with our setup for this computational experiment. We initiate infection with four infected cells (green) placed randomly on the left edge of a tube of cells (blue) with a specified aspect ratio \addednew{($C/L$)} and viral diffusion coefficient $D$.  We conduct simulations for varying choices of aspect ratio and $D$, where the number of cells in the overall tissue is held fixed. (c) Snapshots of the time evolution of a typical infection in this experimental setup with aspect ratio 0.25 and $D=100~\text{CD}^2\text{h}^{-1}$. We show dead cells in black. (d) Time series of the cumulative infected cell proportion for varying $D$ (here aspect ratio is fixed at 0.25). We show ten trajectories in each case. $t_{0.95}^D$ denotes the average time at which simulations with viral diffusion coefficient $D$ infect 95\% of the tissue. (e) Sensitivity to diffusion. For a given tissue aspect ratio, we define diffusion sensitivity as the ratio between time taken to infect 95\% of the tissue with zero diffusion, $t_{0.95}^0$, (or $D = 100~\text{CD}^2\text{h}^{-1}$, $t_{0.95}^{100}$) compared to the 95\% infection time with infinite diffusion, $t_{0.95}^{\infty}$. We show diffusion sensitivity as a function of tissue aspect ratio.}
		\label{fig:diffusion_matters}
	\end{figure*}

	In the literature on spatial viral dynamics, an assumption is sometimes made that the extracellular viral density is approximately uniform across the tissue, or, equivalently, that free viral transport is sufficiently fast as to be effectively instantaneous \citep{blahut_et_al_hepatitis_c_two_modes_of_spread, williams_et_al_inference, goyal_and_murray_CCT_in_HBV}. We tested whether, for a given model cell population size and diffusion coefficient, this assumption would be equally applicable to tube-shaped tissue geometries as it is to more typical square geometries. We began by constructing model tissues in a tubular geometry with a fixed number of cells but varying aspect ratio. Throughout this experiment, we use tissues of $4096$ cells with tube circumference of $C$ cells and a length of $L$ cells such that the aspect ratio of the tissue is $C/L$. Note that aspect ratio is defined based on cell numbers in each dimension and not the actual size of the tissue. Here, and throughout this work, we define tissues in tube geometry as rectangular grids of hexagonal cells with periodic boundary conditions on the upper and lower edges of the sheet and no--flux boundary conditions on the left and right ends of the tube. For this experiment, we initiate infection with four infected cells randomly placed on the left edge of the tissue. We illustrate this setup and present simulations in Figure \ref{fig:diffusion_matters}. Figure \ref{fig:diffusion_matters}(a) shows the topology of the tube, and Figure \ref{fig:diffusion_matters}(b) shows the unrolled representation of the tube with the experimental setup. For full details of the implementation, refer to Section \ref{sec:methods} and the Supplementary Information.

	In Figure \ref{fig:diffusion_matters}(c) we plot tissue snapshots from a representative simulation, in this case with an aspect ratio of 0.25 and diffusion coefficient $D=100~\text{CD}^2\text{h}^{-1}$ (where CD is a cell diameter, here taken to be 10$\mu$m as the size of a typical epithelial cell in the respiratory tract \citep{devalia_et_al_nasal_bronchial_cells}). Figure \ref{fig:diffusion_matters}(c) shows typical dynamics of the model. Infection begins on the left end of the tube. It is initially sluggish, but eventually accelerates and invades deeper into the tube. These snapshots show the formation of a wide infection front moving from left to right, followed by a region of dead cells. We observe the formation of distinct, tightly clustered infection foci beyond the infection front. These distal foci are caused by extracellular viral diffusion away from the infected front, which predominantly spreads via cell--to--cell infection in accordance with our parameter selections.
	
	In Figure \ref{fig:diffusion_matters}(d), we plot the cumulative proportion of infected cells in the tissue as a time series for a range of choices of the extracellular virus diffusion coefficient $D$, ranging from zero to infinity. Tissue aspect ratio is fixed here at 0.25, and we plot ten trajectories for each value of $D$. Figure \ref{fig:diffusion_matters}(d) shows cumulative infected proportion growing approximately linearly with time, but with a dramatic dependence on the value of $D$. Simulations with zero diffusion (i.e., where infection can only progress through cell--to--cell infection, shown in dark green) and those where diffusion is infinite (i.e., where viral concentration is uniform, yellow) exhibit very different time scales of infection. Intermediate values of diffusion give rise to relatively evenly spaced trajectories between these extremes.
	
	\added{To investigate} sensitivity to diffusion for a given choice of tissue aspect ratio, we compute the ratio between the time taken to (almost) completely infect the tissue with no diffusion and the time taken with infinite diffusion. We use the notation $t_{0.95}^D$ for the mean time taken to infect 95\% of the cell sheet when the diffusion coefficient is $D$. We report the time until 95\% infection instead of total infection due to the high degree of noise associated with the latter. In Figure \ref{fig:diffusion_matters}(e) we plot $t_{0.95}^{0}/t_{0.95}^{\infty}$ for varying tissue aspect ratios in orange. \added{Figure \ref{fig:diffusion_matters}(e) shows that infinite viral diffusion results in a faster 95\% infection time than in simulations with finite or zero viral diffusion, but that the extent to which these infection times differ is highly dependent on the tissue aspect ratio. We found that if the circumference and the length of the tube were equal (\addednew{an aspect ratio of 1}), the infection time was increased around five--fold in the zero diffusion case compared to the infinite diffusion case, whereas when the aspect ratio was much smaller (\addednew{1/256}), the zero diffusion case took 80 times longer. This is because, on narrow tubes, infections which can only spread cell--to--cell (that is, with zero viral diffusion) are very slow due to the limited number of susceptible cells neighbouring the infection front at a given time. The system is therefore particularly sensitive to the presence of viral diffusion, since cell--free infections away from the infection front --- even if rare --- dramatically increase access to neighbouring susceptible cells. This effect is much less pronounced on wide tubes, where there is less limitation on susceptible cells neighbouring the infection front.}
	
	\added{In Figure \ref{fig:diffusion_matters}(e),} we also compare the ratio of the time to 95\% infection for the $D=100~\text{CD}^2\text{h}^{-1}$ case --- which we use in the remainder of this work --- relative to the infinite diffusion case, and show these results on the same axes in purple. While this curve shows \added{infection times for the infinite diffusion case are in} closer agreement with the {$D=100~\text{CD}^2\text{h}^{-1}$ case than the} zero diffusion case, there is nonetheless a more than six--fold difference in infection time in the most elongated geometry, and even when the aspect ratio is 0.25, the $D=100~\text{CD}^2\text{h}^{-1}$ case is still around 25\% slower than the infinite diffusion case. Hence, even with this fairly large diffusion coefficient, we still obtain significantly different dynamics compared to when diffusion is infinite. Notably, the change in dynamics \added{observed in Figure~\ref{fig:diffusion_matters}(e)} is a direct consequence of the interplay between viral diffusion and tissue geometry, and not due to a change in the prevalence of the cell--to--cell versus cell--free infection. In Supplementary Figure S2, we plot the mean proportion of infections arising from the cell--to--cell mechanism across ten simulations for a range of viral diffusion coefficients and tissue aspect ratios. Supplementary Figure S2 shows that there is no discernable difference in the balance of the two modes of transmission for any $D\ge10~\text{CD}^2\text{h}^{-1}$ in any given tissue aspect ratio, and moreover that there was negligible variation in this balance \emph{between} tissue aspect ratios relative to the noise of the system. Collectively, Figure~\ref{fig:diffusion_matters}(e) shows that, \addednew{on increasingly narrow tube-shaped tissues (that is, tubes with small aspect ratio),} the dynamics of viral infections \addednew{become increasingly dependent} on not just the presence, but also the rate of viral diffusion. \addednew{These results suggest that} the assumption that viral diffusion can be ignored and approximated as infinite --- \addednew{as is sometimes applied in models which use} square tissue geometries \citep{blahut_et_al_hepatitis_c_two_modes_of_spread, williams_et_al_inference, goyal_and_murray_CCT_in_HBV} --- \addednew{may require greater scrutiny in tissues where infection primarily spreads along one axis and is constrained along another, as is the case for respiratory infections \emph{in vivo}.}

	\subsection{ \added{Tissue} 
		branching structure imposes directionality of infection}

\begin{figure*}[h!]
		\centering
		
		\includegraphics[width=0.92\textwidth, page=3, trim={2.5cm, 10.2cm, 2.5cm, 2.5cm}, clip]{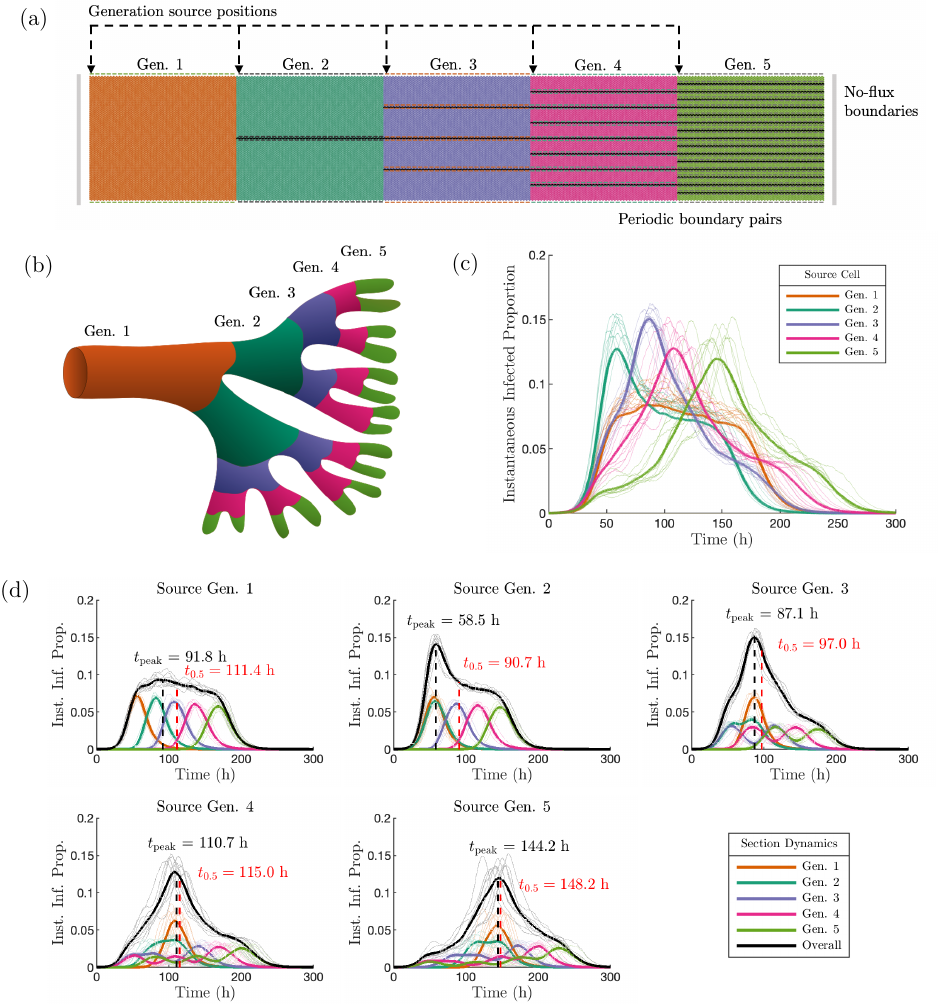}
		
		\caption{(a) An unrolled representation of a branching tree geometry with five generations of branching. Within each generation, branching topology is implemented through pairs of periodic boundaries on the top and bottom of each branch of the generation. We apply no--flux boundary conditions on the left and right edges of the overall tree. We also indicate the left edges of each generation, where we (randomly) place initially infected cells in the following analysis. Here, and throughout this work, we use the same colour scheme for the five tissue generations of the branching tree. (b) Topology of the tissue in (a). (c) Instantaneous infected cell dynamics of infections on a branching domain for infections seeded on the left edge of each tissue generation (ten replicates in each). The tissue is of width 64 cells and length 500 cells, where the tissue branches at every 100 cells along its length. (d) Dynamics (ten replicates and the mean) in each tissue generation for the above trajectories. We also indicate the mean time to peak infected proportion.}
		\label{fig:branching_seed_positions}
	\end{figure*}

	To investigate the role that the branching structure of the \added{respiratory tract} 
	imposes on viral dynamics, we constructed a simple tissue geometry which serves as an approximation of a segment of the airway tree, and perform simulations which are presented in Figure \ref{fig:branching_seed_positions}. An illustration of this construction is shown in Figure \ref{fig:branching_seed_positions}(b). Figure \ref{fig:branching_seed_positions}(b) shows, from left to right, a single tube of tissue which splits into two offspring tubes at each of a sequence of branching points. The regions of tissue between branching points are the generations of the branching tree. We will make the simplifying assumption throughout this work that each tube division is even and binary --- that is, if the circumference of tubes in Generation $g$ is $C_g$, then the circumference of tubes in Generation $g+1$ is $C_g/2$ --- and moreover, that each generation is of equal length. We briefly note that, in reality, measurements of the anatomy of the human respiratory tree suggest that airways in subsequent generations tend to only shrink by around 20\%, and as such, the sum of circumferences in subsequent generations increases substantially \citep{makevnina_lung_atlas, chen_sars_cov_2_in_lung}. In choosing to keep this quantity fixed, our model geometry should therefore be considered a caricature of the structure of the \added{respiratory tract}, 
	instead of an exact representation of the biological reality. This approach, however, enables us to focus our results specifically on the presence of branching structure within the model tissue, and the presence of wider and narrower regions of tissue --- both of which certainly are found in the \added{respiratory tract} 
	--- without the conflating effects of a domain which grows in overall width from left to right. Moreover, this geometry allows for the airway tree to be represented as a two-dimensional grid of cells with judiciously placed periodic boundaries, ensuring computational tractability and avoiding mathematical complications near generation splits. We show a schematic of the unrolled branching tissue in Figure \ref{fig:branching_seed_positions}(a), which shows the pairs of periodic boundaries lining either side of each branch of tissue within a given generation, defining the structure of the branching tree. Throughout this work, we model five tissue generations of length 100 cells each, where the circumference of the Generation 1 tube is 64 cells and the Generation 5 tubes therefore have circumference 4 cells \added{(again, in reality, branches of the human respiratory tract are typically wider in absolute circumference; we use this smaller size to maintain computational tractability while qualitatively representing the \emph{in vivo} structure)}. As with the tube, we place no--flux boundary conditions on the left and right edges of the branching tree. For full details of implementation, refer to Section \ref{sec:methods} and the Supplementary Information.
	
	We simulated viral infection of the branching tree by initiating infection with a single infected cell, where the seed cell was placed randomly on the left edge of a given tissue generation. We carried out ten simulations for each source generation and show the resulting infected proportion time series in Figure \ref{fig:branching_seed_positions}(c). Figure \ref{fig:branching_seed_positions}(c) shows the time series of each individual simulation along with the mean trajectory (excluding simulations where the infection quickly died out). This plot shows that, with the exception of the infection seeded in Generation 1, the dynamics form a clear peak of infection, and are generally delayed the deeper into the tree the infection is seeded. In the Generation 1 case, by contrast, the infection appears to be delayed compared to the Generation 2 case, and moreover forms a plateau in infected cell proportion with no clear peak of infection.
	
	To probe these observations further, for each of the simulations in Figure \ref{fig:branching_seed_positions}(c) we plotted the time series for infected proportion in each tissue generation, along with the overall time series, which we show in Figure \ref{fig:branching_seed_positions}(d). To track how the infection peaks seen in Figure \ref{fig:branching_seed_positions}(c) is affected by source generation, we indicate the mean time of the infection peak at each source generation, for which we write $t_{\text{\tiny peak}}$. As an alternative metric of the time course of infection, we also show the mean time to 50\% infection, $t_{0.5}$. Figure \ref{fig:branching_seed_positions}(d) confirms the observation made above that the peak of infection is delayed the deeper into the branching tree the infection is seeded. Furthermore, it shows that the simulations seeded in Generation 1 have the special property that the individual time series for each tissue generation are tightly peaked and non-overlapping, meaning the overall infected proportion time series is therefore fairly flat. In each other case, the dynamics in individual tissue generations form multiple peaks and overlap substantially. However, in each of these cases, the orange curve --- representing the infected proportion dynamics in Generation 1 --- always forms a single peak which correlates with the overall infected proportion peak. Taken together, these observations reveal that local availability of target cells governs the rate of infection spread. Since the spread of infection is spatially limited by the diffusion of extracellular virus and cell--to--cell infection, infection is faster in regions of the branching tree with the shortest path to the largest number of susceptible cells, which is Generation 1. Here, the infection spreads at its fastest and reaches its peak. Infection spread is likewise initially sluggish when seeded in deeper tissue generations, where the narrowness of the tube restricts the number of susceptible cells within a given distance of the infection. Infections do not reach a single peak in deeper tissue generations, since the infection must travel substantial distances to spread between the branches of a given tissue generation. The upper-- and lowermost branches of Generation 5, for example, are only connected via the top of the tree in Generation 1. Infections seeded in Generation 1, however, have the unique property that, as the infection front spreads down the branching tree, it reaches all the branches of each tissue generation more or less simultaneously. As such, there is little difference in the dynamics of infection in each tissue generation and the overall dynamics are relatively flat.
	
\begin{figure*}[h!]
		\centering
		
		\includegraphics[width=0.7\textwidth, page=4, trim={3cm, 22.5cm, 3cm, 2.5cm}, clip]{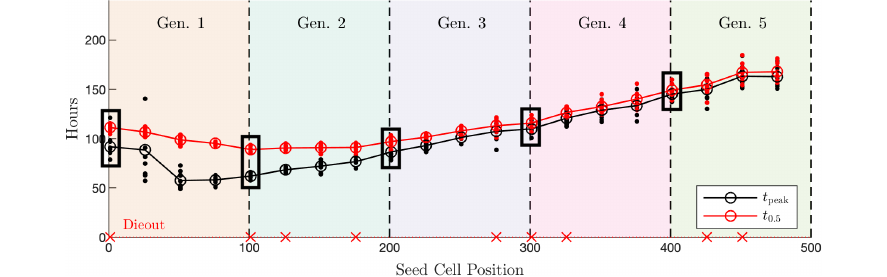}
		
		\caption{Time of peak infected population and time of 50\% total infection for simulations on the branching domain where the sole initially infected cell is placed at different depths along the tissue. We show 10 replicates for each and also indicate the means. Boxed cases were the simulations presented in Figure \ref{fig:branching_seed_positions}(c).}
		\label{fig:branching_seed_fine_res}
	\end{figure*}

	This finding explains the shape and position of the peak of the infected proportion time series, but does not account for why infections seeded on the left edge of Generation 1 should be slower than those seeded in Generation 2. To explore this in more depth, we ran more simulations where the single initially infected cell was placed at different depths (counting from left to right) along the branching tree in finer resolution. We report the time of the peak infected proportion along with the time to 50\% infection for the ten simulations in each case along with the means, and plot our results in Figure \ref{fig:branching_seed_fine_res}. Figure \ref{fig:branching_seed_fine_res} confirms an overall increasing trend in both the time to the infected peak and the time to 50\% infection as the depth of the seed cell increases from Generation 2 onwards, however, it also shows that infections seeded further left of this reach a peak significantly later. In this case, the infection is seeded close to the top of the tree where we have imposed a no--flux boundary. The infection front, therefore, can effectively only spread in one direction (down the tree), slowing the rate of infection. We note that while estimates of the time to 50\% infection are very consistent between iterations of the model regardless of the seed cell position, the time of the infection peak is subject to considerable noise when seeded close to the top of the tree. This is because in these cases as the infection front progresses down the tree, there is little variation in the target cell availability near the front, hence the infected proportion time series is flat and the peak is subject to substantial noise. These two metrics also substantially diverge for seed cell positions in the first two tissue generations, indicating that in these cases, the infected proportion time series is significantly biased. Here, the infection reaches a peak early (as the infection front spreads through Generation 1), then slows down as it spreads through the remainder of the branching tree, reaching 50\% infection of the tissue later.

	\subsection{Tissue geometry and aspect ratio influence the fate of competing viral lineages}\label{sec:extinction}
	
	Recent genetic surveys of the within-host viral populations of individuals infected with influenza have shown a remarkably low degree of genetic diversity \citep{mccrone_et_al_influenza_genetics_in_host}. The same study also found that single nucleotide variants --- the majority of which were synonymous --- from early samples were very likely to be absent from samples taken later in infection. These findings suggest that genetic diversity of the within--host influenza virus population is rapidly lost, and dominated by stochastic rather than evolutionary processes. We sought to use our model to explore the dynamics of competing viral lineages within the tissue. We did so by introducing multiple, colour--coded viral populations. We used this approach --- which was inspired by experimental work by Fukuyama and colleagues, who studied mice co--infected with a suite of differently-coloured influenza reporter viruses \citep{fukuyama_et_al_color_flu} --- in an earlier work, as it has the additional benefit of aiding in visualising the spread of infection from multiple foci \citep{williams_et_al_inference}. This method was implemented by seeding infection with several cells, each infected with viruses of different lineages, and determining the lineage associated with each newly-infected cell by tracking the viral and cell population associated with each lineage. All lineages were assumed to follow the same mechanics and have the same parameters. For full details of implementation, refer to Section \ref{sec:methods}.
	
\begin{figure*}[h!]
		
		\centering
		
		\includegraphics[width=0.82\textwidth, page=5, trim={2.5cm, 7.8cm, 2.5cm, 2.5cm}, clip]{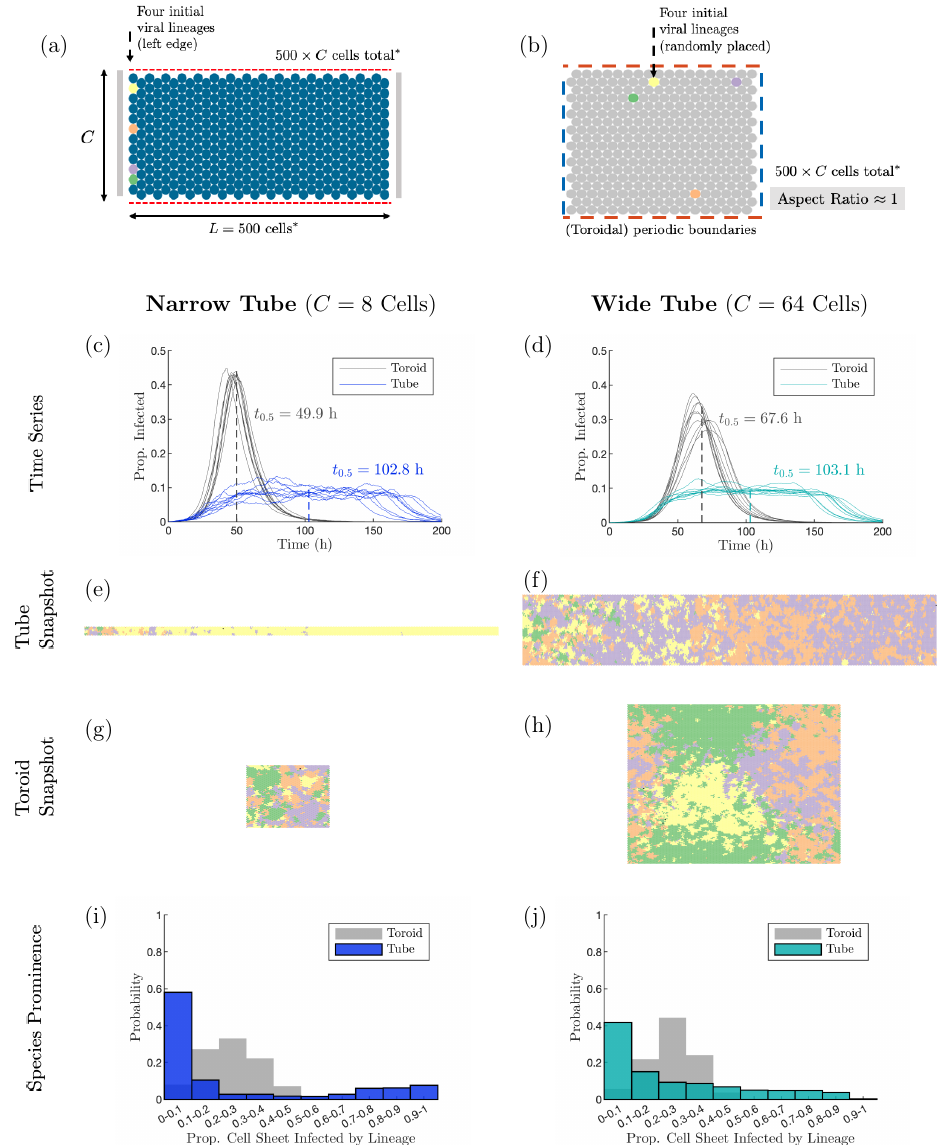}
		
		\caption{(a)--(b) Schematic illustrating experimental setup. We initiate infections with four infected cells of distinct viral lineages on (a) the left edge of a tube of tissue of circumference $C$ cells and length 500 cells, and (b) randomly among an approximately square tissue containing the same number of cells, with toroidal periodic boundary conditions (pairs of dashed lines represent periodicity of boundaries). Infection lineage is computed at each new infection event. Infected cells are coloured by lineage. $^*$These figures are illustrative and are not the same size as tissues used in simulations. (c)--(d) Proportion of cells infected over time on tube and toroid geometry. We also indicate the time to 50\% infection. (e)--(f) Snapshot of tube tissue at the completion of infection, with each cell coloured by the lineage which infected it. (g)--(h) As with (e)--(f), for the toroidal tissue. (i)--(j) Histograms of the proportion of the cell sheet infected by each lineage for the tube-shaped tissue (coloured) and the toroidal tissue (grey).}
		\label{fig:tube_extinction}
		
	\end{figure*}

	As a starting point, we sought to compare the lineage dynamics on tubes of varying circumferences. We did so by constructing tubes of length 500 cells and different choices of circumference $C$. As in an earlier result, we ran simulations of our model by seeding infection with four initially infected cells randomly placed on the left edge of the tube, with the exception that these cells are now assumed to be infected by different viral lineages. As a control, we also compared each case with a ``naive'' approach, where we constructed an approximately square tissue with the same number of cells, and applied toroidal periodic boundary conditions. Note that we define a tissue as ``square'' by having equally many cells in width as in length, however, due to the hexagonal packing of the cells, the physical dimensions of the sheet are rectangular. In the absence of tissue edges in this case, the four seed cells were placed randomly in the toroid. This control case is representative of the standard modelling approach in the literature where the geometry of the \added{respiratory tract} 
	is not considered. Schematics and simulations on different geometries are presented in Figure \ref{fig:tube_extinction}. We show schematics of the tube and toroid geometries in Figure \ref{fig:tube_extinction}(a)--(b). We compare the results obtained for a narrow tube (with a circumference of 8 cells) to those obtained for a wide tube (circumference of 64 cells) --- along with their corresponding toroids --- in Figure \ref{fig:tube_extinction}(c)--(j).
	
	In Figure \ref{fig:tube_extinction}(c)--(d), we plot the infected proportion time series for ten simulations on the narrow and wide tube and their corresponding toroids. We also indicate the time taken to reach 50\% infection of the sheet. Note that we use this metric as opposed to the time to infection peak due to the very flat and noisy trajectories of the tube time series. Figure \ref{fig:tube_extinction}(c)--(d) show that, compared to the toroid, infection of the tube is substantially slower, and the proportion of cells infected at any given time is significantly lower. Also, while infection of the larger toroid (corresponding to the $C=64$ tube) is much slower than for the smaller one (corresponding to the $C=8$ tube) --- owing to the proportionally smaller initial infected population --- the time scale of infection on the narrow and wide tubes is very similar (within 1\%). This is despite the fact that there are eight times more cells in the wide tube case. This finding suggests that the rate of infection along the tube depends primarily on the length of the tube, and that spread along the circumference of the tube is comparatively fast (at least for circumferences up to 64 cells).
	
	In order to visualise the spatial structure of infected cell populations of different viral lineages, we took snapshots of the final state of the tissue for different simulations and coloured each cell by the viral lineage that infected it. In Figure \ref{fig:tube_extinction}(e)--(f) we show representative snapshots of the unrolled tube tissue for the narrow and wide tubes respectively, and in Figure \ref{fig:tube_extinction}(g)--(h) we show representative snapshots of the corresponding toroids. Collectively, Figure \ref{fig:tube_extinction}(e)--(h) show that, while there is qualitatively little difference in the structure of infected populations in the toroid instances, there is a sharp difference between the two tube snapshots. In the snapshot of the narrow tube, the lineages are well-mixed near the left edge of the tube (where infection was seeded) but the orange and green populations rapidly disappear, and the purple cells also become more sparse and are also lost approximately halfway down the tube, leaving only the yellow population. In the wide tube case, however, while the green lineage is only found in the first third of the tube, all other lineages are found throughout the tube, albeit in unequal proportions. By comparison, in both toroid cases, all four viral lineages coexist in relatively similar amounts.
	
	In order to quantify these observations, we ran 100 simulations on both the narrow and wide tube and their corresponding toroids and reported the distribution of proportions of the tissue infected by each lineage. We show these results as histograms in Figure \ref{fig:tube_extinction}(i)--(j). Figure \ref{fig:tube_extinction}(i)--(j) shows two key results. Firstly, the distributions for the toroids and tubes are qualitatively completely different. For the tubes, lineages are by far most likely to infect <10\% of the tissue, with higher proportions markedly less likely, demonstrating dominance by a single lineage (or a small number of lineages). By contrast, for the toroids, density is centred at 20--30\%, corresponding to a relatively even split of the tissue between the four lineages. This distribution is more tightly peaked for the larger toroid due to the larger cell population. This finding suggests that taking the typical approach to multicellular modelling, that is, in taking the cell sheet to be approximately square and with toroidal boundaries, is insufficient for capturing the lineage dynamics seen on 
	tissue geometry \added{more realistic to the respiratory tract}. Figure \ref{fig:tube_extinction}(i)--(j) also show a small but not insignificant probability of a single lineage dominating the cell population (>80\%) in the narrower tube. This is not observed for the wide tube.

	\subsection{Branching of 
		\added{the respiratory tract} promotes loss of 
		viral diversity \added{within airway branches} and spatial isolation of viral lineages}
	
	Having established the results in the previous section for a single tube, we next sought to explore how viral lineage dynamics were impacted by the inclusion of branching structure in the tissue geometry. To do so, we first explored how the lineage dynamics on the branching tree were impacted based on whether the infection started in the upper or lower branches. We conducted a set of 100 simulations on a branching tree where, as with the tubes, infection was initiated with four infected cells of different lineages, randomly placed on the left, open edge of the tree (left edge of Generation 1). Then, for comparison, we also carried out 100 simulations where the seed cells were randomly placed within a single branch on the right, branching edge of the tree (right edge of Generation 5). These two scenarios are chosen to represent, in an abstract way, infections of the upper and lower respiratory tract, respectively. As in Section \ref{sec:extinction}, we again use a branching tree with maximum circumference of 64 cells and five generations of length 100 cells, such that the overall dimensions of the tissue are the same as that of the wide tube, as previously defined. Simulations on tube and branching geometries and analyses are shown in Figure \ref{fig:branching_extinction}. \added{In Supplementary Videos S1 and S2 we show example simulations of infections on the branching geometry from the open edge or the branching edge, respectively.}
	
	We constructed a histogram of the proportion of the tissue infected by each lineage across many simulations, shown in Figure \ref{fig:branching_extinction}(a). Figure \ref{fig:branching_extinction}(a) shows a marked difference in the outcomes of viral lineages for the two different source scenarios. While the distribution for the case where the seed cells are placed on the open edge qualitatively resembles that of the wide tube in Figure \ref{fig:tube_extinction}(j), the distribution for the case where infection is seeded on the branching edge resembles a far more extreme version of that of the narrow tube (Figure \ref{fig:tube_extinction}(i)). When infection is seeded on the branching edge, lineages are fated to either infect virtually none (0--10\%) or virtually all (90--100\%) of the tissue.

	\begin{figure*}[h!]
		\centering
		
		\includegraphics[width=0.9\textwidth, page=6, trim={2.5cm, 16.5cm, 2.5cm, 2.5cm}, clip]{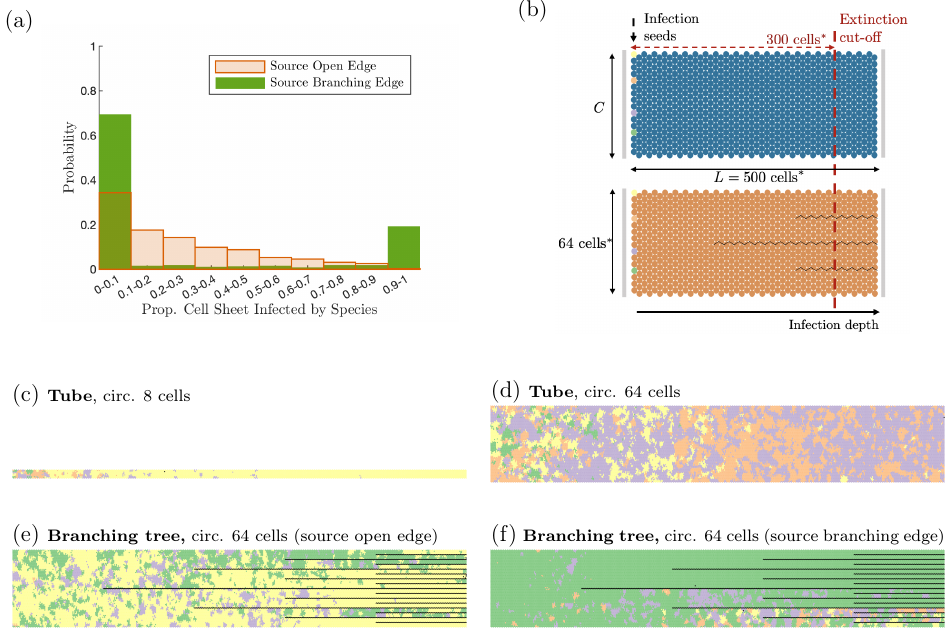}
		
		\caption{Distribution of infection lineages under different tissue topology. In each case, we seed infection with four cells of different lineages and attribute to each new infection the lineage responsible. We assign to each infection lineage a different colour. (a) Histogram of the proportion of the cell sheet infected by each lineage for infections of the branching tree seeded on the open edge of the tree (left edge of Gen. 1) vs the branching edge (right edge of Gen. 5). (b) Schematic illustrating our definition of an extinction event, where a lineage is not found beyond a cut-off depth along the domain. Here we use a cut-off depth of 300 cells. $^*$These figures are illustrative and are not the same size as tissues used in simulations. (c)--(f) Snapshots of tissues of varying geometry following infection, with cells coloured by the lineage which infected it.}
		
		\label{fig:branching_extinction}
	\end{figure*}
	
	In Figure \ref{fig:branching_extinction}(e)--(f) we show representative snapshots of the unrolled branching tissue for both cases of the source cell positions, along with the tube snapshots from the previous section for comparison (Figure \ref{fig:branching_extinction}(c)--(d)). Figure \ref{fig:branching_extinction}(b) gives a schematic of these simulations. Figure \ref{fig:branching_extinction}(e)--(f) provide clear visual evidence for the sharp contrast between the lineage dynamics of infection on the branching tree starting in the upper or lower branches. These snapshots also suggest that the tissue is more likely to be dominated by a single lineage in the branching tree case (seeded on the branching edge) compared to the narrow tube case due to the larger overall cell population. As can be seen in Figure \ref{fig:branching_extinction}(f), by the time the infection spreads the length of the tree and reaches Generation 1, there is essentially only one lineage left but the entire upper half of the tree remaining to be infected, resulting in that lineage infecting a far greater overall proportion of the total cell population. \added{The lineage dynamics, including crowding--driven extinction events, can be seen emerging over time in Supplementary Videos S1 and S2.} 
	
	As another descriptive tool for analysing the lineage dynamics of infections on varying tissue geometries, we next developed a notion of lineage ``extinction'' if a lineage is not found beyond some specified cut-off depth into the tissue (we use a cut-off of 300 cells). We denote by $\pext$ the probability that, in a given geometry, a given lineage will go extinct. Figure \ref{fig:branching_extinction}(b) shows a schematic of this construction. Clearly, extinction as we have described it relies on a single direction of viral invasion, and is therefore not well defined for infection of the branching tube from the branched edge. In the narrow tube, we computed $\pext=0.633$, meaning that a given lineage was more likely than not to go extinct, whereas for the wide tube, we found $\pext=0.245$, meaning extinction only occurred around 25\% of the time. Interestingly, on the branching tree (with source on the open edge), the probability of extinction was slightly \emph{reduced} at $\pext=0.200$, despite containing the same number of cells as the wide tube but with the inclusion of narrow branches within its geometry. This finding results from the fact that the branching structure acts to keep lineages separate, which slightly reduces the chance of their being crowded out. This can be seen in the lineage structure at the right end of both the wide tube and the branching tube (with source on the open edge). While in the tube case the remaining lineages are well mixed across the tube's circumference, in the branching case, each of the narrow terminal branches is almost entirely uniform in lineage. When infection is seeded in these narrow branches, however, the branching structure does not act to separate and protect lineages since, as was observed in the narrow tube (Figure \ref{fig:branching_extinction}(c)), lineages are rapidly crowded out before they have a chance to be separated and lineage diversity is quickly lost. \added{In brief, within branching tissues, there is a tradeoff between spatial restriction imposed by the narrowing of the branches, which \emph{promotes} lineage extinction, and the physical separation of different branches, which prevents the populations of the branches of a given generation from competing with each other, thereby \emph{reducing} the chance of lineage extinction. We found here that the overall effect of the branching structure is to slightly decrease lineage extinction overall compared to a tissue of the same size without branching (the wide tube geometry).}
	
\begin{figure*}
		
		\centering
		
		\includegraphics[width=0.82\textwidth, page=7, trim={2.5cm, 10cm, 2.5cm, 10cm}, clip]{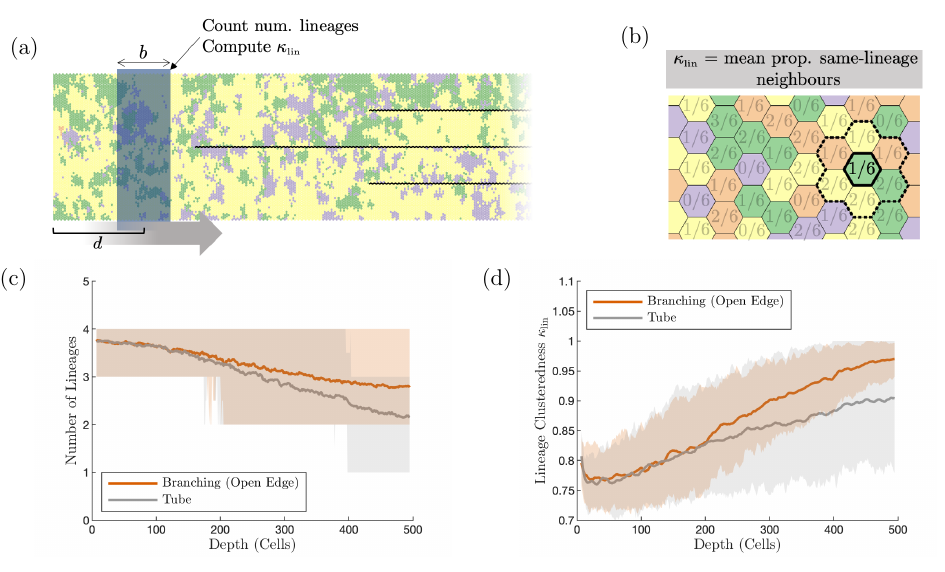}
		
		\caption{(a) Schematic of analysis in this Figure. We construct a moving band of cells of bandwidth $b$ centred on depth $d$ cells along a tissue. Within the band we compute the number of lineages and the clustering metric $\kappalin$. (b) Definition of the clustering metric $\kappalin$, which, given a tissue in which each cell has been assigned a lineage, is the mean proportion of a cell's neighbours which share the same lineage. (c)--(d) Number of lineages and lineage clusteredness, $\kappalin$, at varying tissue depths, respectively. We compute values for a band of cells of width 10 cells, centred on the indicated depth. In each case, we compare results for the wide tube and the branching tube, seeded on the open edge (that is, (d) and (e) in Figure \ref{fig:branching_extinction}). We show the mean along with the 10th and 90th percentiles.}
		
		\label{fig:branching_band_analysis}
		
	\end{figure*}

	To further quantify the difference in lineage structure on the branching tree (seeded on the open edge) and the wide tube, we analysed properties of the lineage dynamics along the depth of the tissue. To do so, for a given depth $d$ cells along the tissue, we examined the band of cells with depth $d-b/2$ to $d+b/2$ (where $b$ is the bandwidth, taken to be 10 cells). We sketch this computation process in Figure \ref{fig:branching_band_analysis}(a). For each band of cells analysed, we computed both the number of lineages found in the band, and a clustering metric, $\kappalin$, as a measure of the extent to which lineages are clustered together. This metric, which is similar to one which we developed elsewhere \citep{williams_et_al_inference}, is computed on a cell population after an infection is complete and simply reports the mean proportion of a cell's neighbours which are of the same lineage as the cell. We show a schematic of this calculation in Figure \ref{fig:branching_band_analysis}(b). We plot both the number of lineages and the clustering metric $\kappalin$ as a function of band depth $d$ in Figure \ref{fig:branching_band_analysis}(c)--(d). In both cases we indicate the mean along with the 10th and 90th percentiles. Figure \ref{fig:branching_band_analysis}(c)--(d) both show that for the first approximately 200 cells of depth, the two geometries are relatively indistinguishable, and the difference in the final number of lineages remaining is not substantial compared to the noise of the system. However, for $\kappalin$, we observe that, as well as there being a notable divergence in the means for the branching tree and tube case, there is also a substantial difference in the amount of variation in each case. While $\kappalin$ is an extremely noisy metric towards the right end of the tissue, it is very consistently close to 1 for the branching tube. This confirms our observation from the snapshots that while lineages are somewhat mixed in the tube case, the branches in Generation 5 of the branching geometry generally contain only a single lineage. Note that the reason for the small range of $\kappalin$ values (around [0.7,1]) is a result of the high proportion of infections arising from cell--to--cell infection, leading to a high probability of same-lineage neighbour cells. For a different ratio of cell--to--cell to cell-free infections, it is possible that the difference in $\kappalin$ trajectories for the branching tree and tube geometries would be more pronounced.

	\subsection{Immune response depends on the location of infection within a branching structure}

	Our simulations so far have not explicitly considered the role of the immune response. Resolution of infection in our model is controlled only by the availability of the susceptible cell population, a defining feature of target cell--limited models. However, actual respiratory infections \emph{in vivo} do not result in the destruction of the entire 
	\added{respiratory} epithelium. Instead, only reasonably localised regions are affected, such as the upper airways or the alveolar passages, before the host immune response or drug intervention results in the clearing of infection \citep{gallagher_spatial_spread}. We sought to investigate the behaviour of a basic model of immune activity, and how it might interact with infections seeded in different regions of the branching tree geometry.
	
\begin{figure*}[h!]
		\centering
		
		\includegraphics[width=0.86\textwidth, page=8, trim={2.5cm, 16cm, 2.5cm, 2.5cm}, clip]{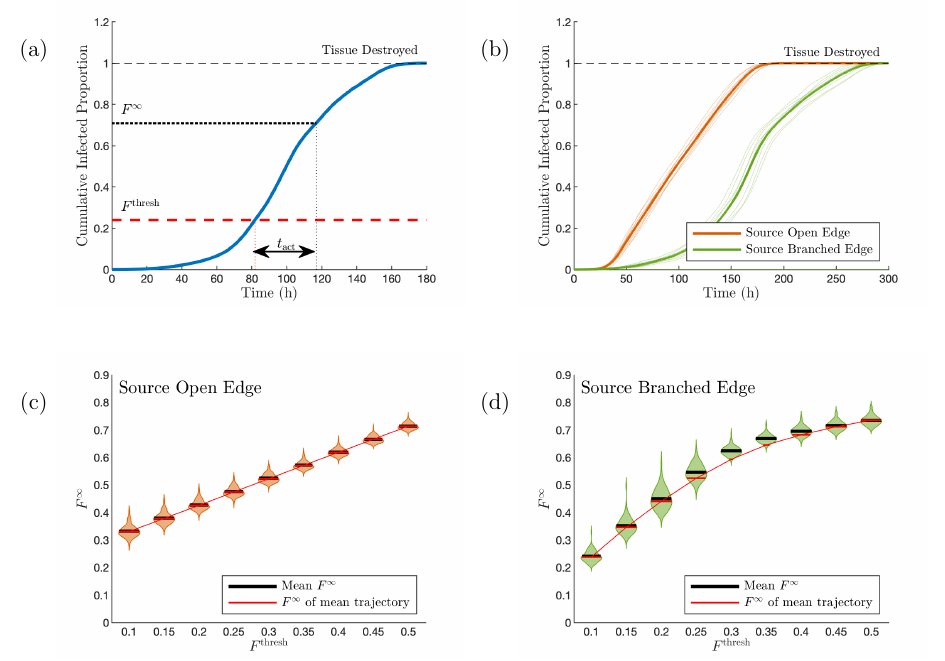}
		
		\caption{(a) Schematic of a simple model for the immune response in the \added{respiratory tract}. 
			We first specify a threshold, $\Fthresh$ for cumulative infected cell proportion. Once the infection meets the threshold, we assume the immune response is triggered, and then takes $t_{\text{act}}$ hours to activate. Once activated, we assume that the immune response immediately clears the infection. We report the cumulative infected proportion of the tissue at this point, $\Finf$, as a measure of damage to the tissue. (b) Time series for the cumulative infected proportion for infections on the branching tree, where infection is initiated with a randomly placed single cell on either the left edge of Gen. 1 (the open edge of the geometry), or on the right edge of Gen. 5 (the branched edge). In each case, we plot ten trajectories and the mean, along with an indicative value of $\Fthresh$. (c)--(d) Cumulative tissue damage on the branching tree after immune action for varying $\Fthresh$ and fixed $\tact=30$h, where the infection is seeded on the open edge or the branched edge, respectively. We show violin plots for $\Fthresh$ across 100 simulations (mean in black). We also plot in red the predicted final cumulative damage where the immune response is applied to the mean of the trajectories.}
		\label{fig:immune_seed_pos}
		
	\end{figure*}

	We introduce the following simple model for the immune response, which is presented and investigated in Figure \ref{fig:immune_seed_pos}. We assume some threshold proportion, $\Fthresh$, such that, once the cumulative infected proportion of the tissue reaches $\Fthresh$, the host immune response is triggered. At this point, there is a delay of $\tact$ hours while the immune response activates, then, once the response is active, we assume that the infection is immediately and totally cleared. At this point, we compute the final cumulative infected proportion $\Finf$ as a measure of the cumulative damage to the tissue. We show a schematic of this construction in Figure \ref{fig:immune_seed_pos}(a).

	As in the previous result, we sought to apply our simple model of the immune response to infections of the branching tree beginning from either the open left edge of Generation 1, or from the highly branched right edge of Generation 5. We recall that these two scenarios are chosen to represent upper-- and lower--respiratory infections \emph{in vivo}, respectively. In each scenario, as in Figure \ref{fig:branching_seed_positions}, we initiate infection with a single infected cell randomly placed on the specified edge of the tissue. We plot representative behaviour of both infection scenarios in Figure \ref{fig:immune_seed_pos}(b). Figure \ref{fig:immune_seed_pos}(b) shows the cumulative infected proportion over time for infections seeded on either the open or the branched edge. In each case we plot ten representative trajectories and the mean trajectory. The dynamics of infections progressing from either end of the tissue geometry are substantially different from each other. For infections initiated on the open edge, the cumulative infected proportion grows roughly linearly throughout, whereas for infections spreading from the branched edge, initial growth is slow and exponential, followed by a change in convexity at a cumulative infected proportion of around 0.7. \added{Figure \ref{fig:immune_seed_pos}(b) shows that the trajectories of individual simulations are qualitatively similar whether the source is on the open or branched edge of the tissue.}

	We began by testing the final cumulative damage to the tissue, $\Finf$, for each of these infection seed positions for varying values of the detection threshold, $\Fthresh$. We ran 100 simulations of the model for infections of the branching tree seeded either on the open or the branched edge, and applied our immune model for a range of $\Fthresh$ values between 0.1 and 0.5. Here, we have kept $\tact$ fixed at 30h. We constructed violin plots for the distribution of $\Finf$ values at each $\Fthresh$ value, and annotated these with the mean $\Finf$ value. We compared this to the $\Finf$ computed on the mean \emph{trajectory} for both source positions, which we also show on the same axes. That is, we compare the mean of the $\Finf$ on the individual trajectories and compare this to the $\Finf$ on the mean trajectory. We show these results for both the open and branched edge source in Figure \ref{fig:immune_seed_pos}(c) and (d), respectively. Figure \ref{fig:immune_seed_pos}(c)--(d) show that cumulative damage increases roughly linearly on this range of detection thresholds for the open edge case, while for the branched edge case, damage increases at a sublinear rate. This corresponds to the different curvature in the cumulative infected curves for the two scenarios up to around 70\% infected. Infections seeded on the branched edge are also subject to greater variation in the final damage to the tissue between iterations of the model, especially when $\Fthresh$ takes a value around 0.2. Finally, we also note that while the mean $\Finf$ and the $\Finf$ of the mean trajectory agree well for the open edge source case, where the overall variation in final cumulative damage is minimal, there is a notable difference between the two for the branched edge source case. Here, especially for $\Fthresh\in[0.2,0.4]$, the mean trajectory sustains a substantially lower final proportion of damage compared to the average damage in the individual runs \added{(the $\Finf$ computed on the mean trajectory has a $z$-score of -1.86 compared to the $\Finf$ on individual trajectories at $\Fthresh=0.3$ and -1.56 at $\Fthresh=0.35$; for infections seeded on the open edge of the tissue, $z$-scores do not exceed 0.14 in magnitude across the range of $\Finf$ values tested)}. This difference in $\Finf$ suggests that stochastic effects in individual runs of the model, \added{at least for some initial conditions or perhaps in some tissue geometries, may} have a substantial effect on the outcome of the immune response 
	and that averaged, deterministic models (as others have used in spatial viral dynamics \citep{bocharov_et_al_dynamics_virus_infection_space_time, bocharov_et_al_spatiotemporal_dynamics_virus}) may not fully capture the observed dynamics.

\begin{figure*}[h!]
		\centering
		
		\includegraphics[width=0.86\textwidth, page=9, trim={2.5cm, 14.7cm, 2.5cm, 2.5cm}, clip]{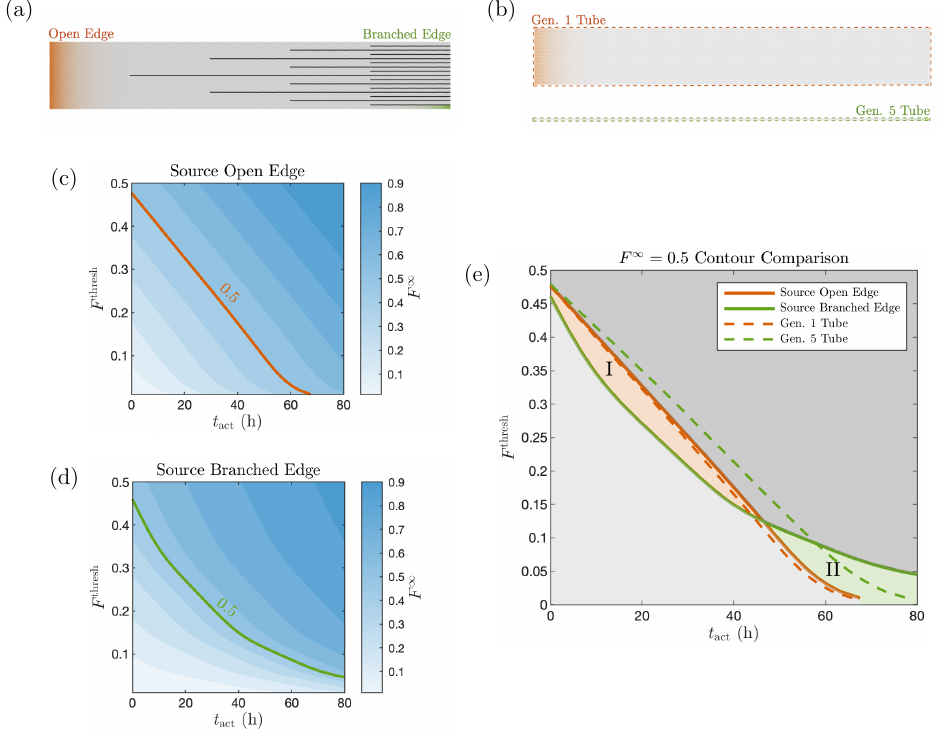}
		
		\caption{(a) Schematic illustrating experimental setup. We initiate infection on a branching domain with a single infected cell either on the open edge of the tree (left edge of Gen. 1) or the branched edge (right edge of Gen. 5). (b) For reference, we also compare these simulations against infections on tubular domains of equal length, and circumference equal to either the widest tissue generation (Gen. 1) or the narrowest tissue generation (Gen. 5) of the branching tree. (c)--(d) Contour plots of the cumulative damage to the tissue, $F^{\infty}$, following immune action for varying $t_{\text{act}}$ and $F^{\text{\tiny thresh}}$, where the immune response is defined as in the previous figure. We show results for infections on the branching tube where infection is seeded on the open or branched edge of the tree, and highlight the $F^{\infty}=0.5$ contours. (e) Comparison of the $F^{\infty}=0.5$ contours for infections on the branching domain, where infection is initiated on the open or branched edge of the tree. We also plot the $F^{\infty}=0.5$ contours for the wide and narrow tube of equal length to the branching tree. In regions I and II of parameter space, the immune response more effectively targets infections seeded on the open edge, and the branched edge, respectively.}
		
		\label{fig:immune_param_sweep}
		
	\end{figure*}

	Next, we decided to examine how the interplay of the two immune parameters, $\Fthresh$ and $\tact$, would influence the final cumulative damage to the two infection seed positions. To do this, we computed the final infected proportion, $\Finf$, on each of the 100 simulations mentioned above over a wide range of values for $\Fthresh$ and $\tact$. As a control, we applied the same method to infections on tubes the same length as the branching tree and with the same circumference as either the widest branch (64 cells) or the narrowest branch (4 cells) of the tree. These tissues represent a geometry where the tube that the infection started in continues out to the length of the full branching tree, but without any of the branching structure (a tube of circumference either 64 cells on the open edge, or 4 cells on the branched edge). We show a schematic of this setup in Figure \ref{fig:immune_param_sweep}(a)--(b).

	In Figure \ref{fig:immune_param_sweep}(c)--(d) we show a heatmap of $\Finf$ values on a range of values for $\Fthresh$ and $\tact$. We also display the $\Finf =0.5$ contour in both cases, which we use as a rough threshold for whether the host survives the infection. Figure \ref{fig:immune_param_sweep}(c)--(d) show that, predictably, when the immune response is both sensitive and quick to activate, very little damage is sustained, whereas when the immune response requires a higher threshold of detection and more time to become active, the tissue is mostly destroyed. The two cases differ in the shape of the contours between these extremes.
	
	In order to compare these two scenarios, we plotted the $\Finf=0.5$ contours from both cases on the same axes in Figure \ref{fig:immune_param_sweep}(e). We also include the $\Finf=0.5$ contours on the same region of $(\tact,\Fthresh)$ parameter space for the wide and narrow tubes as defined above (that is, tubes of the same length as the branching tree and circumference equal to that of the widest or narrowest branch of the tree respectively). Figure \ref{fig:immune_param_sweep}(e) shows that the curves for the two source positions on the branching tree intersect, enclosing two distinct regions. This overlap is not a special property of the $\Finf=0.5$ contour; other contours exhibit similar patterns (see Figure \ref{fig:immune_param_sweep} and Supplementary Figure S3). In Region I of Figure \ref{fig:immune_param_sweep}(e), the immune response is triggered only at a high threshold, but activates quickly. In this case, infections seeded in Generation 1 (upper respiratory) are more effectively cleared by the immune response. In Region II, where the immune detection threshold is very low but there is a substantial lag in its activation, infections seeded in Generation 5 (lower respiratory) are more effectively targeted. In the curves for the two control cases --- the wide and narrow tubes --- we do not observe this overlap, indicating that this is a unique property of the branching tree. Our results indicate that even with an extremely simple immune response, we see heterogeneous responses to infections in the upper and lower regions of the branching tree based on the parameters of the immune response.

	\section{Discussion}

	
	In this work, we have explored how the characteristic geometry of the \added{respiratory tract} 
	can influence how 
	viruses like influenza and SARS--CoV--2 spread through 
	tissue. Here, we extended a standard multicellular \changed{computational} model of viral infection to account for two key features of \added{respiratory tract} 
	geometry --- the tubular structure of airways, and the branching process between airway generations --- and studied the role these considerations play in the dynamics of the model.

	Our work showed that imposing this \changed{realistic} geometry on the model tissue results in the emergence of key aspects of infection dynamics which do not arise on \changed{flat tissues such as \emph{in vitro} systems}. For one, we showed that when a cell population of fixed size is arranged in a tubular geometry of a given aspect ratio, the model dynamics are more sensitive to the value of the extracellular viral diffusion coefficient when the tube is narrower and longer. This \addednew{suggests} that \addednew{the rate of viral} diffusion is more important in \addednew{infection dynamics on tissues which are spatially constrained along their circumference relative to their length --- as in the respiratory tract ---} than it is on \changed{flat, square tissues}.
	
	We also observed that, for infections of a tissue structured as a branching tree, the dynamics of infection were influenced by the site of the source of infection. We found that infections seeded in deeper, more highly--branched regions of the tree took longer to infect the tissue, and discovered that this was a property of the abundance of target cells local to the infection front. At least for the parameter values used in this work, we observed infections forming well--defined fronts of infection. As such, on the branching tree, the point at which infection progressed fastest --- and the point of the highest infected cell load --- was when the infection front reached the unbranched edge of the tree (the ``top'' of the tree) where the number of susceptible cells near the front was maximised. Lower on the tree, the branching structure meant that infection fronts were more constrained to only spread along one dimension and therefore had less access to nearby target cells. \changed{Our observation that viral infection spreads more quickly when seeded in wider, upper airways than when seeded in the lower airways is a hypothesis which could be assessed \emph{in vivo}.}

	We showed that \changed{models of viral dynamics which incorporate} realistic tissue geometry for the \added{respiratory tract} 
	\changed{--- including the computational model we have discussed here ---} offer insights into important biological questions. We studied the fate of competing viral lineages within the tissue to explore the role that tissue structure may play in fostering or hindering the spread of viral mutants within the host. Using our model, we observed complex lineage dynamics, and found that, if infection is seeded concurrently by multiple competing viral lineages on the edge of a tube of cells, the likelihood that at least one lineage would be rendered extinct by stochastic effects was greatly increased over the same length of tube when the tube was narrow. This might indicate a mechanism for the very low viral genetic diversity observed within individuals infected with influenza, for example \citep{mccrone_et_al_influenza_genetics_in_host}. The stochastic extinction of lineages on narrow, crowded geometries is a known phenomenon in the ecological literature, called an ``embolism effect'' \citep{bialozyt_embolism_effect, fayard_long_distance_dispersal}. A very similar extinction effect is also well--studied in cellular systems such the colorectal crypt or cancer growth under the name ``neutral drift'' \citep{romijn_et_al_crypt_neutral_drift, huels_et_al_cancer_neutral_drift}. We have shown the potential presence of this behaviour in infections of the \added{respiratory tract}.
	We note that \changed{recovering} such dynamics is only possible in a model \changed{system} that considers tube geometry explicitly: a naive ``control case'' using a more typical computational tissue for the same experiment failed to capture relevant competition dynamics. Interestingly, we found that the branching structure complicates this behaviour, with \added{the inclusion of a branching process in the tissue improving the chance of multiple viral lineages coexisting in the tissue, by spatially separating competing lineages}. \added{Significantly, these complex extinction dynamics are a consequence of the specific structure of the branching tissue geometry, and cannot be simply explained by standard target cell limitation arguments alone. In a biological context,} this result hints at complexities in the way in which viral lineages \added{spread} and are compartmentalised within the \added{respiratory tract}. 
	\added{This process may also be relevant} to the emergence of viral variants within the host for influenza or SARS--CoV--2. \added{Experimental work to probe the spatial distribution of viral genetic variants within the respiratory tract \emph{in vivo} (for example, spatial imaging or sequencing methods) coupled with \emph{in silico} models capable of describing the structure of the respiratory tract, as introduced here, would provide insight into these complex dynamics.}

	We also saw that infections of a branching tissue interacted very differently with a simple immune response model based on whether infection progressed from the open, unbranched end, or from the deeper, highly branched end, of the tree. Importantly, we also observed that when the immune response was slow to detect the infection but had a rapid activation time, it more efficiently cleared infections seeded on the open end of the tree, whereas when the immune response was highly sensitive but sluggish to activate, it more efficiently targeted infections seeded on the branched edge of the tree. We showed that this transition in infection outcomes was specifically a property of the branching geometry: on tubular tissues, wider tissues always sustained more damage following clearance of infection. The infection sources on the open and branched edges of the branching tree are analogous to upper and lower respiratory infections, which are well known to result in widely different host immune responses and pathogenicity \citep{ke_perelson_et_al_covid_upper_and_lower_rt, wolfel_et_al_virological_assessment_covid, gallagher_spatial_spread, taubenberger_and_morens_lung_influenza_pathology}. Our results suggest that geometry alone may provide an explanation for a difference in immune interaction with the two infections scenarios, and moreover indicates that \changed{model systems} which incorporate branching tissue geometry \changed{are necessary to account for} the different disease outcomes associated with infection in these locations.

	\changed{The collection of experimental data is necessary to empirically assess the findings of our simulation study, however, there are substantial challenges in doing so. Perhaps most significantly, our work suggests that standard \emph{in vitro} systems for respiratory viral infections, on flat, structurally heterogeneous tissues, lack a description of important features of infection spread within \added{the} host \added{respiratory tract}. 
		A more complex experimental setup such as an \emph{in vivo} system may therefore be necessary in order to obtain relevant data. To our knowledge, there is no such data currently available in the literature, and the generation of such data is beyond the scope of the present study, but we sketch here possible approaches that could be taken in future work. To assess the variable rates of infection progression at different depths within the respiratory tract, a comparison could be made between the extent of infection in experimental organisms over time --- either via viral titre, or through imaging --- where the initial viral challenge is delivered via different means. The different modes of viral delivery would need to be configured such that the initial sites of viral deposition are at different depths within the \added{respiratory tract}. 
		The study of competing viral lineages within the \added{respiratory tract} 
		has clearer experimental precedent. Fukuyama \emph{et al.} developed a suite of murine influenza strains expressing differently--coloured fluorescent proteins, collectively dubbed ``color--flu'' \cite{fukuyama_et_al_color_flu}. While imaging of the distribution of the different strains required sacrificing of the infected mice, this nonetheless presents an experimental system from which quantitative data can be collected about the spatial distribution of lineages within the \added{respiratory tract}. 
		\jmochanged{Following our previous work on fitting computational models to data, \cite{williams_et_al_inference}, metrics} such as the lineage clustering metric, $\kappalin$, which we have employed here, will facilitate the integration of such data with computational models.}

	
	There are some important limitations to the approach we have applied here. Clearly the model of the respiratory tree we have presented is highly idealised, and future work will consider greater biological realism. For instance, the distribution of cell types in regions of the respiratory tree are varied and therefore so is the distribution of cellular receptors specific to a given virus, meaning that uniform infectivity is not observed throughout the \added{respiratory tract} 
	\citep{gallagher_spatial_spread, gizuarson_mucus_velocity_ciliated_cells, fiege_et_al_covid_tropism_single_cell}. Moreover, in this work we have considered extracellular viral transport to be isotropic and ignored the effects of the mucociliary escalator or the movement of air in the respiratory lumen. The properties of these flows have been shown to vary along the depth of the respiratory tree and likely influence the dynamics of spreading viral infections 
	\citep{gizuarson_mucus_velocity_ciliated_cells, asgharin_mucus_velocity_by_airway}. We anticipate that these heterogeneities will be of greater importance over the entire 24 branching generations of the respiratory tree as opposed to the only five considered in this work \citep{makevnina_lung_atlas}.

	We have also in this work presented only a simple sketch of immune activity. Future work will improve the realism of this mechanism, especially in considering its inherently spatial mode of action: imaging of the lungs of individuals with influenza or SARS--CoV--2 infection reveal diffuse regions of inflammation and infection alongside focal lesions of of concentrated immune activity at sites of infection \citep{camp_et_al_neutrophil_recruitment, taubenberger_and_morens_lung_influenza_pathology}. \added{There is a substantial literature of more detailed, spatially--explicit models for the immune response in viral infections which account for, for example, an interferon--mediated innate immune response \citep{lavigne_et_al_interferon_signalling_ring_vaccination, segredo_otero_et_al_spatial_structure_innnate_immunity}, or the T cell recruitment and migration process \citep{sego_et_al_covid_model, levin_et_al_T_cell_search_influenza}. These kinds of immune processes will need to be included in future versions of the model in order to improve the faithfulness of the model to biological reality.} Nonetheless, \added{even the very basic model of the immune response which we use here is sufficient to} 
	show that the geometry of the tissue alone is sufficient to lead to complex and varied interactions, 
	suggesting that this may play a role in the different disease outcomes for upper and lower respiratory infections.
	
	\added{Our model neglects a number of aspects of virus--host interactions --- heterogeneity of cell types, anisotropic spread of viruses, details of the immune response, for example --- in the interest of honing our focus on the effect of the tissue geometry on the dynamics. In order to be used to make robust and quantitative inferences from experimental data, we would expect that at least some of these aspects would need to be incorporated into the model. Deciding which of these extensions would be necessary to interpret data from a given experimental system would likely need to be a bespoke process led by model selection and inference--based methods.}
	
	Finally, we acknowledge that future work will need to accommodate biologically realistic measurements for the actual sizes and structures of respiratory tissues \emph{in vivo}. For instance, in our work we consider airway branches with circumferences as small as 4 cells, while in reality, even the narrowest human airways are likely to be at least 30--130 cells in circumference \citep{chen_sars_cov_2_in_lung, makevnina_lung_atlas} (assuming respiratory epithelial cells have a diameter of around $10\mu m$ \citep{devalia_et_al_nasal_bronchial_cells}). In order to study tissues with elongated aspect ratios while limiting the overall cell population to a computationally tractable size, it was necessary to consider tubes with circumferences below what is biologically realistic. \addednew{As a consequence, while the tissues simulated here are sufficient models to provide qualitative insights into how viral dynamics differ on tubular or branching structures (compared to flat tissues, for example), it is likely that the extent to which certain effects arise in absolute, quantitative terms, will differ on tissues of realistic dimensions. Therefore, in order} to provide a more detailed, quantitative insights into \emph{in vivo} dynamics, larger model tissues of accurate dimensions will be needed.
	
	\added{Moreover, in} our construction of a branching tree geometry, we have made the strong assumption that tissue generations are of equal length and that branching is an equal, binary process, such that the sum of circumferences of the branches of a given tissue generation is fixed. We have already discussed how this does not reflect the actual dimensions of respiratory trees \emph{in vivo}. In reality, the sum of circumferences of consecutive generations of the \added{respiratory tract} 
	tend to expand to fill space, especially in the terminal airways \citep{makevnina_lung_atlas}. \added{Individual airways also tend to be shorter relative to their circumferences than those shown in Figure \ref{fig:diffusion_matters} or Figure \ref{fig:branching_seed_positions}, for example, and this is especially true of the terminal airways.} We show what unrolled tissues would realistically look like, using measurements \addednew{of human airway branches} for the dimensions \citep{makevnina_lung_atlas}, in Supplementary Figure S4. Here we use the same colour scheme for the generations as we have used in the main body of this work. \added{While the tissue geometries we have used in this work are clearly simplifications of the \emph{in vivo} reality, they are still qualitatively useful models of the respiratory tract structure which capture the main features of interest, in particular, the branching process of the airways and the spatial confinement within branches.} \added{To integrate our findings with \emph{in vivo} applications, an investigation of model dynamics on} a more complex and biologically accurate tissue geometry is an important avenue for future work, \added{and is not a trivial extension of the current study}.

	We have explored how explicitly including the geometry of the respiratory tree \changed{in} \jmochanged{computational} models of 
	viral infections \changed{reveals rich dynamics of infection which are not well--described by conventional approaches (both computational and experimental) \jmochanged{which only consider} flat, square model tissues}. \added{This fact may have broader consequences for infections of the respiratory tract more generally, including infections by other pathogens such as bacteria.} \changed{Model systems which incorporate a more realistic geometry} are also able to account for the complex dynamics of competition between viral lineages within the host, which might offer an important tool for studying the emergence and extinction of genetic variants. We moreover showed that infection spread in upper and lower regions of the branching tree generate distinct dynamics. This suggests that the difference in tissue geometry alone may influence the varying dynamics of upper and lower respiratory infections.

	\section*{Declaration of competing interest}
	
	The authors declare no competing interests.

	\section*{Code availability}
	
	All code is freely available on the GitHub repository \url{https://github.com/thomaswilliams23/lung_structure_virus_dynamics}.

	\section*{Acknowledgments}
	
	TW's research is is supported by an Australian Government Research Training Program (RTP) scholarship. JMM's research is support by the ARC (DP210101920). JMO's research is supported by the ARC (DP230100380, FT230100352). The funders had no role in study design, data collection and analysis, decision to publish, or preparation of the manuscript.

	\appendix
	
	\section{Supplementary Information}
	
	Supplementary material related to this article can be found in the accompanying ``Supplementary Information'' document.

	\clearpage
	\bibliographystyle{apalike}
		
\end{document}


\maketitle

	\section{A simple multicellular model for viral dynamics}
	
	We assume that the computational domain, $\Omega \subset \mathbb{R}^2$, comprises $N$ discrete, spatially explicit cells, indexed by $i=1, 2, ..., N$. Cell $i$ is assumed to occupy the fixed spatial region $S_i$. We associate with each cell $i$ at time $t$ a specific state, denoted by $\sigma_i(t)$, where \mbox{$\sigma_i(t)\in\left\{T,E,I,D\right\}$}, representing the susceptible (\textbf{T}arget) state; infected, but not yet infectious (\textbf{E}clipse) state; productively \textbf{I}nfectious state; or \textbf{D}ead state. 
	
	We also track the extracellular viral density $v(\mathbf{x}, t)$ on $\mathbf{x}\in\Omega$. As in our previous study, we assume that extracellular virus is secreted by productively infectious cells, however, unlike in our previous work, we also assume here that extracellular virus is a spatially variable quantity. We assume that extracellular virus is secreted uniformly over the spatial regions occupied by productively infectious cells at a rate $p$, that virus density diffuses across the tissue according to linear diffusion with coefficient $D$, and that its density also decays uniformly at rate $c$. Collectively, $v(\mathbf{x},t)$ is governed by the PDE
	\begin{equation} \label{eq:virus_pde}
	\pd{v}{t} = p\sum_{i \in \mathcal{I}(t)} \frac{ \mathbbm{1}_{\left\{\mathbf{x} \in S_i\right\}}}{\left| S_i \right|} - cv +D \nabla^2 v,
	\end{equation}
	
	\noindent where $\mathcal{I}(t) = \left\{i~:~\sigma_i(t)=I \right\}$ is the set of productively infectious cells at time $t$, and where $\mathbbm{1}_{\left\{\cdot\right\}}$ is the usual characteristic function. This is a standard model of extracellular viral spatial dynamics in the literature (e.g., \cite{williams_et_al_spatial_discretisation, sego_et_al_covid_model}).
	
	Transitions between cell states are determined probabilistically, and are assumed to follow a Poisson process. We assume that ``susceptible'' to ``latently infected'' transitions --- that is, infection events --- are mediated by two mechanisms: infection by extracellular virus, and infection via cell--to--cell contact. In extracellular virus infection, we assume that a cell $i$ can only be infected by only the virus in the region of space which it occupies, that is, $\int_{S_i}v(t)d\mathbf{x}$. We assume this mode of infection is controlled by rate $\beta$. For cell--to--cell infection, we assume that the rate of infection is dependent on the proportion of a cell's neighbours which are productively infectious at a given time. That is, if we denote by $\nu(i)$ the set of indices of the cells neighbouring cell $i$, and by $n_{\text{\tiny neighbours}}=6$ the fixed number of neighbours a cell can have, the probability of cell $i$ becoming infected by cell--to--cell infection depends on the term $\sum_{j\in\nu(i)} (\mathbbm{1}_{\left\{\sigma_j(t)=I\right\}})/n_{\text{\tiny neighbours}}$. We assume that this mode of infection is controlled by the rate $\alpha$.
	
	Collectively, we arrive at the probability of infection over the small time interval $[t, t+\dt)$:
	\begin{equation}\label{eq:T_to_E}
	P(\sigma_i(t+\dt)=E \vert \sigma_i(t)=T) = 1-\exp\left(-\left(\alpha\sum_{j\in\nu(i)}\frac{ \mathbbm{1}_{\left\{\sigma_j(t)=I\right\}}}{n_{\text{\tiny neighbours}}} + \beta N \int_{S_i} vd\mathbf{x}\right)\dt\right).
	\end{equation}
	
	\noindent Note that the inclusion of the $N$ next to $\beta$ ensures dynamics remain in agreement with an ODE form of the model. We performed this calculation in an earlier work \cite{williams_et_al_spatial_discretisation}.
	
	For the ``latently infected'' to ``productively infectious'' transition, we assume, following others, that the duration of the eclipse phase obeys a gamma distribution with parameters $K$ and $1/K\gamma$ \cite{petrie_et_al_reducing_uncertainty_influenza_method_of_stages, williams_et_al_inference, fain_and_dobrovolny_gpu_data_fitting}. That is, if we write $t_i^E = \min\{t : \sigma_i(t) =E\}$ for the time at which cell $i$ enters the latently infected state, and $t_i^I = \min\{t : \sigma_i(t) =I\}$ for the time at which cell $i$ enters the productively infected state, we have
	\begin{equation}
	t_i^{I} = t_i^{E} + \tau_i,
	\end{equation}
	
	\noindent where
	\begin{equation}
	\tau_i\sim \text{\emph{Gamma}}\left(K,\frac{1}{K\gamma}\right).
	\end{equation}
	
	\noindent This can be thought of as introducing $K$ latently infected sub-states into the model, each of which has an exponentially distributed duration. The mean total time spent in the latently infected state is therefore $1/\gamma$.
	
	The ``productively infectious'' to ``dead'' state transition is assumed to be governed only by a fixed death rate, $\delta$, such that the probability of a productively infectious cell dying in the time interval $[t,t+\dt)$ is given by
	\begin{equation}\label{eq:I_to_D}
	P(\sigma_i(t+\dt)=D \vert \sigma_i(t)=I) = 1 - \exp\left(-\delta\dt\right).
	\end{equation}
	
	\noindent Collectively, Equations \eqref{eq:virus_pde}--\eqref{eq:I_to_D} govern the multicellular model. 
	
	Throughout this work, we will assume fixed values of the model parameters as specified in Table \ref{tab:default_params}. These values were selected simply to be indicative of the realistic range of values for these parameters and are sufficiently realistic for the purposes of this work. The values of $K$, $\gamma$, $\delta$, $p$, and $c$ were obtained by running a Bayesian parameter estimation for a simpler version of this model against experimental data for influenza infection published by Kongsomros \emph{et al.}. We then selected one particular posterior sample at random \cite{kongsomros_et_al_trogocytosis_influenza}. We used the same parameter values in an earlier study; refer to this work for further details on their selection \cite{williams_et_al_inference}. 
	
	Moreover, throughout this work, we also assume the fixed values for $\alpha$ and $\beta$ listed in Table \ref{tab:default_params}. These values were selected using a lookup table we also constructed in our previous work \cite{williams_et_al_inference} and ensure that, given the values of the other parameters (and infinite diffusion), toroidal geometry, and a random 1\% of the tissue initially infected, simulations on toroidal geometry should reach a peak infected proportion at approximately $25$h, with approximately 90\% of the infections arsing from the direct cell--to--cell mechanism. The high weight given to cell--to--cell infection is consistent with experimental observations for SARS-CoV-2 \cite{ zeng_et_al_sars_cov_2_cell_to_cell}.  For further details on this construction, refer to our earlier study \cite{williams_et_al_inference}.
	
	Finally, unless otherwise specified, we will assume a value of 100 CD$^2$h$^{-1}$ for the diffusion coefficient $D$ throughout this work. This value was chosen to ensure qualitative agreement between simulated infection spreading patterns and those observed \emph{in vivo}. This value is consistent with estimates of the diffusion of influenza or SARS--CoV--2 virions in water or plasma at body temperature \cite{sego_et_al_covid_model, holder_et_al_design_considerations_influenza, beauchemin_et_al_modeling_influenza_in_tissue}.

	\begin{table}[h!]
		\centering
		
		\begin{tabular}{p{6.5cm}|p{1.5cm}|p{5.8cm}}
			
			\hline
			\textbf{Description} & \textbf{Symbol} & \textbf{Value and Units}  \\
			\hline
			Cell--to--cell infection rate& $\alpha$&$1.839483~\text{h}^{-1}$\\
			Extracellular virus infection rate& $\beta$&$2.694176\times10^{-8}\text{(TCID}_{50}\text{/ml)}^{-1}\text{h}^{-1}$\\
			Number of delay compartments& K& 3\\
			Eclipse cell activation rate& $\gamma$ & $3.366934\times10^{-1}\text{h}^{-1}$ \\
			Death rate of infected cells& $\delta$ & $8.256588\times10^{-2}\text{ h} ^ {-1}$  \\
			Extracellular virion production rate& $p$ & $1.321886\times10^6 (\text{ TCID}_{50}\text{/ml}) \text{ h}^{-1}$  \\
			Extracellular virion clearance rate& $c$ & $4.313531\times10^{-1} \text{ h}^{-1}$\\
			Extracellular viral diffusion coefficient& $D$ & $100~\text{CD}^2\text{h}^{-1}$ (unless specified)\\[0.5em]
			
		\end{tabular}
		\caption{Fixed parameters used in our simulations.}
		\label{tab:default_params}
	\end{table}

	\section{Parameter selection}\label{sec:parameter_selection}
	
	We use a fixed set of model parameters throughout this work. The model parameters used in this work, with the exception of the infection parameters $\alpha$ and $\beta$, were derived in an earlier work by our group \cite{williams_et_al_inference} which used an ODE model with the same structure as the multicellular model used here to fit experimental data for influenza \cite{kongsomros_et_al_trogocytosis_influenza}. Refer to our publication for further details on the fitting process. 
	
	Since the infection parameters $\alpha$ and $\beta$ could not be uniquely identified in the fitting process, we specified them by first constructing a lookup table. We ran repeated simulations of the multicellular model as defined in this work at each point of an array of $\alpha$ and $\beta$ values. For the look-up table we used a 50$\times$50 grid of cells with toroidal boundary conditions and infinite extracellular viral diffusion. Infections were seeded with 1\% of the cells randomly chosen as initially infected. We computed the mean proportion of cell--to--cell infections (as opposed to cell--free infections) as well as the time of peak infected cell proportion at each $\alpha$--$\beta$ combination across the iterations (we used 20). From these lookup tables, we used splines to interpolate between points to determine the $\alpha$ and $\beta$ values that correspond to a given cell--to--cell infection proportion and a given peak time. For this work, we selected the $\alpha$ and $\beta$ values that corresponded to approximately 90\% cell--to--cell infection and a peak time of 25h.
	
	\pagebreak

	\section{Tissue structures}\label{sec:tissue_structures}
	
\begin{figure}[h!]
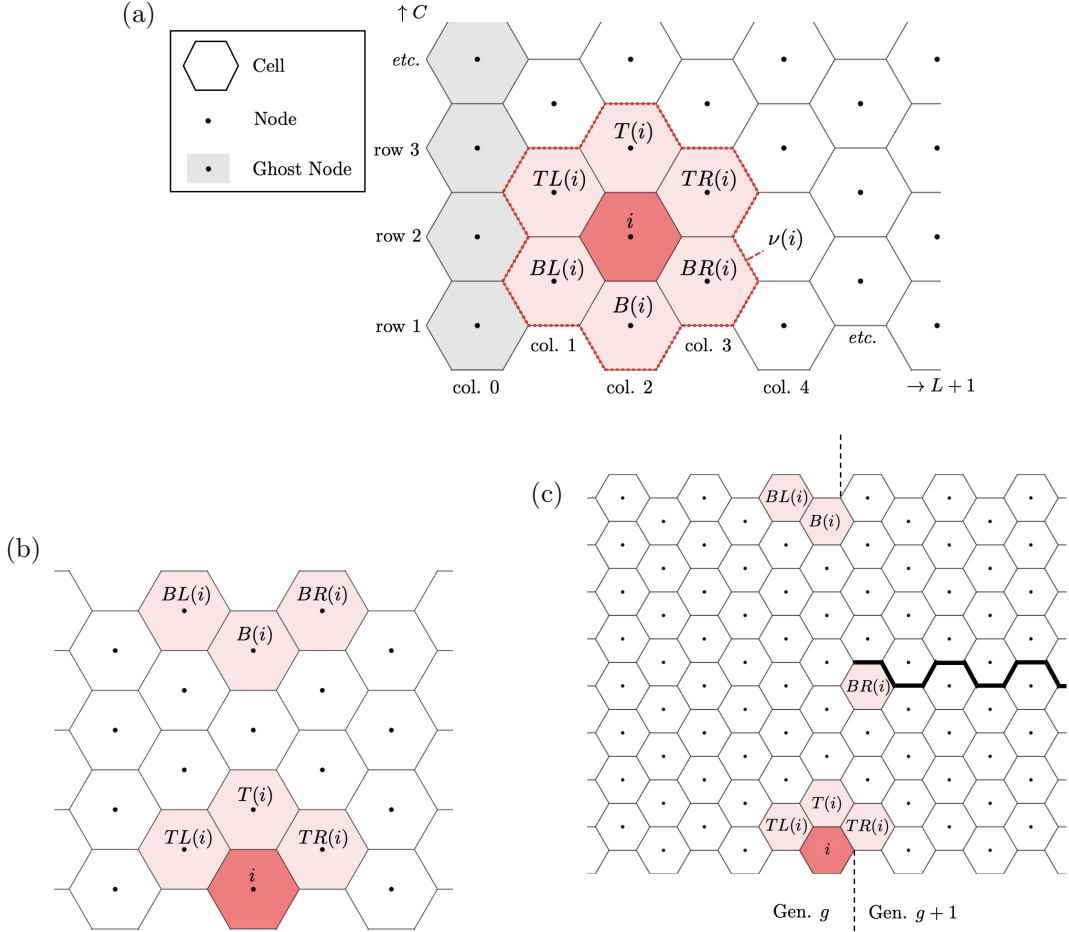

		\centering
		\begin{subfigure}{0.8\textwidth}
			\centering
			\raisebox{14em}{(a)}
			\includegraphics[width=0.9\linewidth]{overall_numerics}
		\end{subfigure}
		
		\vspace{1em}
		
		\begin{subfigure}{0.4\textwidth}
			\raisebox{14em}{(b)}
			\includegraphics[width=0.9\linewidth]{tube_numerics}
		\end{subfigure}
		\hspace{2em}
		\begin{subfigure}{0.5\textwidth}
			\raisebox{16em}{(c)}
			\includegraphics[width=0.9\linewidth]{branching_numerics}
		\end{subfigure}
		\caption{(a) Schematic illustrating the notation and methods of the numerical scheme. (b) Adjacency on tube geometry, (c) adjacency on branching geometry.}
		\label{fig:numerics_cartoon}
	\end{figure}

	Throughout this work, we assume two-dimensional model tissues with hexagonal packing of cells.  Each cell occupies an equally-sized regular hexagonal region of space and, except at boundaries, which we discuss below, each cell has precisely six neighbours. Figure \ref{fig:numerics_cartoon} illustrates this geometry. Hexagonal packing of cells reflects the reality of epithelial cell packing and has the practical benefit that all cell--to--cell contacts occur at an edge, with no complications arising from corner neighbours.
	
	As can be seen in Figure \ref{fig:numerics_cartoon}, the hexagonal packing of the cells naturally results in the formation of distinct columns of cells, with an offset on alternating columns. To refer to the position of a cell, we adopt the notation $(x(i), y(i))$ to define its position, where $x$ is the column index (counting from the left) and $y$ is the row index (counting from the bottom). We also introduce the inverse notation, $i = n(x,y)$ for the index of the cell in column $x$ and row $y$. With this notation, we can explicitly define the cell adjacency function $\nu(i)$ for different geometries.
	
	For \emph{tubes} of circumference $C$ cells and length $L$ cells, we represent the unrolled tissue as a rectangular $L\times C$ sheet of cells with periodic boundary conditions in the $y$ direction and no--flux boundary conditions in the $x$ direction. We must assume that $C$ is even to ensure that the hexagonal grid of cells closes; we moreover will always take $L$ to be even. Then for cell $i$, we have
	\begin{equation}\label{eq:adjacency_general_def}
	\nu(i) = \begin{cases}
	\left\{T(i),B(i),BR(i),TR(i)\right\},\qquad\qquad\qquad~~~ x(i)=1,~\text{(left edge)}\\
	\left\{BL(i),TL(i),B(i),T(i),BR(i),TR(i)\right\},~~~ 1<x(i)<L,~\text{(tube interior)}\\
	\left\{BL(i),TL(i),T(i),B(i)\right\},\qquad\qquad\qquad~~~ x(i)=L,~\text{(right edge)}
	\end{cases}
	\end{equation}
	
	\noindent where $BL(i),~TL(i),~B(i),~T(i),~BR(i),~TR(i)$ are the bottom-left, top-left, bottom, top, bottom-right, and top-right neighbours of cell $i$, respectively. We have for the tube geometry
	\begin{align}
	BL(i) &= \begin{cases}
	n(x(i)-1, R_{\text{\tiny tube}}(y(i)-1)),\qquad~~ x(i)\text{ even,}\\
	n(x(i)-1, y(i)),\qquad\qquad\qquad~~~~ x(i)\text{ odd,}
	\end{cases}\\
	TL(i) &= \begin{cases}
	n(x(i)-1, y(i)),\qquad\qquad\qquad~~~~ x(i)\text{ even,}\\
	n(x(i)-1, R_{\text{\tiny tube}}(y(i)+1)),\qquad~~ x(i)\text{ odd,}
	\end{cases}\\
	B(i) &= n(x(i), R_{\text{\tiny tube}}(y(i)-1)),\\
	T(i) &= n(x(i), R_{\text{\tiny tube}}(y(i)+1)),\\
	BR(i) &= \begin{cases}
	n(x(i)+1, R_{\text{\tiny tube}}(y(i)-1)),\qquad~~ x(i)\text{ even,}\\
	n(x(i)+1, y(i)),\qquad\qquad\qquad~~~~ x(i)\text{ odd,}
	\end{cases}\\
	TR(i) &= \begin{cases}
	n(x(i)+1, y(i)),\qquad\qquad\qquad~~~~ x(i)\text{ even,}\\
	n(x(i)+1, R_{\text{\tiny tube}}(y(i)+1)),\qquad~~ x(i)\text{ odd,}
	\end{cases}
	\end{align}
	
	\noindent where
	\begin{equation*}
	R_{\text{\tiny tube}}(y) = \texttt{mod}(y-1,C)+1
	\end{equation*}
	
	\noindent accounts for the periodic boundaries of the tube, where we use the shorthand
	\begin{equation*}
	\texttt{mod}(a,n) = a\mod n.
	\end{equation*}

	For the \emph{branching} geometry, under the assumption that branching is even and binary, and that the resulting offspring branches are each half the circumference of the preceding tissue branching generation, we can represent the unrolled tissue as a two-dimensional sheet as follows. We assume that branching tree comprises $G$ tissue generations, such that Generation $h=1,2,...,G$ contains $2^{h-1}$ tubes of cells, and that the circumference of the single tube in Generation 1 is $C$ cells. It follows that the circumference of a tube in a given Generation $h$ is $C/2^{h-1}$ cells. We assume that the overall length of the branching tree from the left to right edges is $L$ cells, and that ${b_1, b_2,...,b_G}$ are the depths, in numbers of cells from the left edge, of the first cell of each tissue generation, where $b_1=1$. As before, we assume that the left and right edges of the branching tree are subject to no--flux boundary conditions.
	
	Under this construction, as for the tube geometry, the general definition of the adjacency function $\nu(i)$ in Equation \eqref{eq:adjacency_general_def} remains valid, and we can represent the unrolled branching tree as an $L\times C$ sheet of cells, with the addition of special conditions for cells lying along the boundaries of tube branches. For cell $i$ in a branching tree, we simply adjust the edge reflection terms to obtain the definitions
	\begin{align}
	BL(i) &= \begin{dcases*}
	n\left(x(i)-1, \rminus_{\text{\tiny branch}}\left(x(i)-1,y(i)-1\right)\right),\qquad~~x(i)\text{ even,}\\
	n(x(i)-1, y(i)),\qquad\qquad\qquad\qquad\qquad\qquad~~ x(i)\text{ odd,}
	\end{dcases*}\\[1em]
	TL(i) &= \begin{dcases*}
	n(x(i)-1, y(i)),\qquad\qquad\qquad\qquad\qquad\qquad~~ x(i)\text{ even,}\\
	n\left(x(i)-1, \rplus_{\text{\tiny branch}}\left(x(i)-1,y(i)+1\right)\right),\qquad~~ x(i)\text{ odd,}
	\end{dcases*}\\[1em]
	B(i) &= n\left(x(i),  \rminus_{\text{\tiny branch}}\left(x(i),y(i)-1\right)\right),\\
	T(i) &= n\left(x(i), \rplus_{\text{\tiny branch}}\left(x(i),y(i)+1\right)\right),\\[1em]
	BR(i) &= \begin{dcases*}
	n\left(x(i)+1,  \rminus_{\text{\tiny branch}}\left(x(i)+1,y(i)-1\right)\right),\qquad~~ x(i)\text{ even,}\\
	n(x(i)+1, y(i)),\qquad\qquad\qquad\qquad\qquad\qquad~~ x(i)\text{ odd,}
	\end{dcases*}\\[1em]
	TR(i) &= \begin{dcases*}
	n(x(i)+1, y(i)),\qquad\qquad\qquad\qquad\qquad\qquad~\ x(i)\text{ even,}\\
	n\left(x(i)+1, \rplus_{\text{\tiny branch}}\left(x(i)+1,y(i)+1\right)\right),\qquad~~ x(i)\text{ odd,}
	\end{dcases*}
	\end{align}

	\noindent where 
	\begin{equation*}
	\rminus_{\text{\tiny branch}}(x,y) = \begin{cases}
	y,\qquad \qquad~~~\texttt{mod}(y,c(x))\ne 0,\\
	y+c(x),\qquad \texttt{mod}(y,c(x))=0,
	\end{cases}
	\end{equation*}
	
	\noindent and
	\begin{equation*}
	\rplus_{\text{\tiny branch}}(x,y) = \begin{cases}
	y,\qquad \qquad~~~\texttt{mod}(y,c(x))\ne 1,\\
	y-c(x),\qquad \texttt{mod}(y,c(x))=1,
	\end{cases}
	\end{equation*}
	
	\noindent account for the periodic boundaries of the tubes in a given tissue generation. We use the shorthand
	\begin{equation*}
	c(x)=2^{g(x)-1}
	\end{equation*}
	
	\noindent for the circumference of a tube in column $x$, and
	\begin{equation*}
	g(x) =\max_{h=1,2,...,G}\left\{x\ge b_h\right\},
	\end{equation*}
	
	\noindent for the tissue branching generation of a cell in column $x$.

	\section{Tracking lineages} \label{sec:tracking_lineages}
	
	In certain simulated infections, we wish to track multiple viral lineages simultaneously. To account for this, we use the following procedure, which we developed in an earlier publication \cite{williams_et_al_inference}.
	
	Suppose we wish to track $\Nlin$ viral lineages. In what follows, we assume each viral lineage to possess identical infection parameters, however in principle this method could easily be extended to accommodate lineages with differing properties. We begin by augmenting the system to account for infection states specific to each lineage, and a corresponding viral density. Specifically, for each lineage $l=1,2,...,\Nlin$, we denote the latently infected, productively infectious, and dead sub-states associated with that  lineage as $E_l$, $I_l$, and $D_l$, respectively. Similarly, if by $\sigma_i^{\text{\tiny sub}}(t)$ we denote the \emph{sub}--state of cell $i$ at time $t$,  the viral density of lineage $l$ is given by
	\begin{equation}
	\pd{v_l}{t} = p\sum_{i \in \mathcal{I}_l(t)} \frac{ \mathbbm{1}_{\left\{S_i\right\}}}{\left| S_i \right|} - cv_l +D \nabla^2 v_l,
	\end{equation}
	
	\noindent where $\mathcal{I}_l(t) = \left\{i~:~\sigma_i^{\text{\tiny sub}}(t)=I_l \right\}$. Note that $\sum_{i=1}^\Nlin v_l = v$, and that $\mathcal{I}(t)=\cup_{i=1}^\Nlin \mathcal{I}_l(t)$, and so on.
	
	We determine the lineage of a newly infected cell as follows. Assume, following the infection event probability defined in Equation \eqref{eq:T_to_E}, that cell $i$ is marked to enter the latently infected state during time interval $[t, t+\dt)$. The probability that cell $i$ does \emph{not} enter the $E_l$ sub--state during time interval $[t, t+\dt)$ is given by 
	\begin{align}
	P(\sigma_i^{\text{\tiny sub}}(t+\dt)\ne E_l \vert \sigma_i(t+\dt)=E,~&\sigma_i(t)=T )\nonumber\\&=P(\mathcal{L}_i \ne l)\nonumber\\ &=\exp\left(-\left(\alpha\sum_{j\in\nu(i)}\frac{ \mathbbm{1}_{\left\{\sigma_j(t)=I_l\right\}}}{\left|\nu(i)\right|} + \beta V_l\right) \dt\right),
	\end{align}
	
	\noindent where the more convenient notation $P(\mathcal{L}_i \ne l)$ has the obvious definition. Therefore, we determine the probability that cell $i$ is infected with lineage $l$ as 
	\begin{align}
	P(\mathcal{L}_i = l) =  \frac{1-P(\mathcal{L}_i \ne l ) }{\sum_{m=1}^{\Nlin}\left(1 - P(\mathcal{L}_i \ne m)  \right)}.
	\end{align}
	
	\noindent We then assign viral lineage as follows. First, draw a random number $x \sim~Uniform(0,1)$, then compute
	\begin{equation}
	l^* = \min\left\{l~:~x<\sum_{m=1}^{l}{P(\mathcal{L}_i = m) }\right\},
	\end{equation}
	
	\noindent that is, the minimum $l$ such that the probability of cell $i$ having a lineage of at most $l$ is greater than $x$. Cell $i$ is then assigned lineage $l^*$.

	\section{Numerical methods}
	
	We simulate our model by stepping time in increments of length $\dt$ (throughout this work, we use $\dt=0.01\text{h}$). At a given time $\tau$, we denote by $\mathcal{G}^{\tau}$ the state of the cell grid, and by $v_{\cdot}^\tau$ the discretised form of the viral density surface $v(\cdot, \tau)$. Then, during the time step $[\tau, \tau+\dt)$, we perform the following.
	
	Firstly, following the state transition probabilities in Equations \eqref{eq:T_to_E}--\eqref{eq:I_to_D}, we check each cell for a state transition, and generate the cell grid for the next time step, $\mathcal{G}^{\tau+\dt}$, based on the state of the cell grid and viral density surface at the \emph{start} of the time step, that is, $\mathcal{G}^{\tau}$ and $v_{\cdot}^\tau$, respectively. 
	
	We then update the discretised viral surface using an implicit--explicit finite--difference scheme. We discretise the viral density in space such that the cells themselves may be considered the nodes of the discretised surface. As a consequence, the total viral density at cell $i$ at time $\tau$ is trivially computed as
	\begin{equation*}
	\int_{S_i}v(\mathbf{x},\tau)d\mathbf{x} = v_i^{\tau},
	\end{equation*}
	
	\noindent where $v_i^{\tau}$ is the value of the discretised viral surface at node (cell) $i$. In an earlier work, we discussed the discretisation of such viral surfaces, and found that when diffusion is sufficiently large compared to the length scale of the cell, discretisation at the cell scale was sufficient to ensure convergence of the virus PDE \cite{williams_et_al_spatial_discretisation}. Throughout this work, we employ very rapid diffusion compared to the length scale of the cell (by default, $100~\text{CD}^2\text{h}^{-1}$), which justifies this choice of discretisation.
	
	For the viral diffusion, we use a Backwards--Euler method constructed on the hexagonal lattice of nodes (cells). Since we use no--flux boundary conditions on the left and right edges of the tissue, we introduce $C$ additional ghost nodes on both the left and right edges of the tissue in columns 0 and $L+1$, following the notation introduced above. We extend the definition of the functions $n(x,y)$, $x(i)$ and $y(i)$ to also assign a single index $i=N+1,N+2, ..., N+2C$ to ghost node $g_i$ residing at position $(x,y)$ and vice versa, and define the new adjacency function
	\begin{equation}
	\nu^*(i) = 
	\nu(i)\cup\nu_{\text{\tiny edge}}(i),
	\end{equation}
	
	\noindent on $i=1,2,...,N$, that is, for the cell nodes, where
	\begin{equation*}
	\nu_{\text{\tiny edge}}(i) = \begin{cases}
	\left\{TL(i), BL(i)\right\},\qquad\qquad x(i)=1,~~\text{(left edge)}\\
	\emptyset,\qquad\qquad\qquad\qquad\qquad~1<x(i)<L,\\
	\left\{BR(i), TR(i)\right\},\qquad~~~~~ x(i)=L.~~\text{(right edge)}
	\end{cases}
	\end{equation*}
	
	\noindent Then, on $i=N+1,N+2,...,N+2C$ --- that is, for the ghost nodes --- define
	%
	%
	\begin{equation*}
	\nu_{\text{\tiny ghost}}^{\text{\tiny col}}(i) = \left\{T(i), B(i)\right\},
	\end{equation*}
	
	\noindent and
	\begin{equation*}
	\nu_{\text{\tiny ghost}}^{\text{\tiny int}}(i) = \begin{cases}
	\left\{BR(i), TR(i)\right\},\qquad\qquad x(i)=0,~~\text{(left ghosts)}\\
	\left\{TL(i), BL(i)\right\},\qquad\qquad x(i)=L+1.~~\text{(right ghosts)}
	\end{cases}
	\end{equation*}
	
	\noindent for the neighbours of the ghost nodes lying in the ghost column, and in the interior of the node grid, respectively.
	
	Overall, the scheme for the update step is given by the matrix equation
	\begin{equation}\label{eq:v_ext_update}
	\hat{\mathbf{v}}_{\text{\tiny imp}}^{\tau+\dt} = \mathbf{D}^{-1}\hat{\mathbf{v}}^{\tau},
	\end{equation}
	
	\noindent where 
	\begin{equation}
	\hat{\mathbf{v}}^{\tau} = \left\{\mathbf{v}^\tau, \mathbf{g}^\tau\right\} = \left\{v_1^{\tau}, v_2^{\tau}, ..., v_N^{\tau}, g_{N+1}^{\tau}, g_{N+2}^{\tau},...,g_{N+2C}^{\tau}\right\},
	\end{equation}
	
	
	\noindent and $\mathbf{D}$ is the $(N+2C)\times(N+2C)$ discretised diffusion matrix, which reflects the adjacency structure of the nodes,  such that
	\begin{equation*}
	\mathbf{D}_{i,j} = \begin{dcases}
	(1+\frac{4D\dt}{\dx^2}), \qquad i=j,~j=1,2,...,N+2C,\\
	-\frac{2}{3}\frac{D \dt}{\dx^2},\qquad ~~~~ i\in\nu^*(j),~j=1,2,...,N,\\
	-\frac{2}{3}\frac{D \dt}{\dx^2},\qquad ~~~~ i\in\nu_{\text{\tiny ghost}}^{\text{\tiny col}}(j),~j=N+1,N+2,...,N+2C,\\
	-\frac{4}{3}\frac{D \dt}{\dx^2},\qquad~~~~  i\in\nu_{\text{\tiny ghost}}^{\text{\tiny int}}(j),~j=N+1,N+2,...,N+2C,\\
	0, \qquad \qquad \qquad~~\text{otherwise}.
	\end{dcases}
	\end{equation*}

	\noindent Here, $\dx$ is the distance between cell centres (cell diameter, or CD). Throughout this work, we work in units of CD, and hence take $\dx=1$. In an update step, we compute the value of the discretised virus surface at time $\tau+\dt$ using Equation \eqref{eq:v_ext_update} (in practice, we use sparse system solvers instead of the computationally expensive process of finding the inverse of $\mathbf{D}$). We then set $g_{n(0,y)}^{\tau+\dt}=v_{n(1,y)}^{\tau+\dt}$, $g_{n(L+1,y)}^{\tau+\dt}=v_{n(L,y)}^{\tau+\dt}$ for $y=1,2,...,C$ to account for the no--flux boundary conditions.
	
	Note that this method only really depends on the definition of the cell adjacency function $\nu$, meaning it applies to the definitions for both the tube and branching tree geometries we have described (and in fact more generally). We skirt potential complexities arising from the complex topology of the sheet by essentially approximating the continuous viral diffusion process as a discrete, cell-based process.
	
	Having computed the viral diffusion step, we then apply an explicit scheme for the remaining terms of the virus PDE:
	
	\begin{equation}
	\mathbf{v}_{\text{\tiny exp}}^{\tau+\dt} = \mathbf{v}^{\tau} + \dt\left( p\mathbf{I}^{\tau} - c \mathbf{v}^{\tau}\right),
	\end{equation}
	
	\noindent where 
	
	\begin{equation*}
	\mathbf{I}^{t} = \left\{\frac{ \mathbbm{1}_{\left\{\sigma_1 (t)=I\right\}}}{N\left|S_1\right|}, \frac{\mathbbm{1}_{\left\{\sigma_2 (t)=I\right\}}}{N\left|S_2\right|}, ..., \frac{\mathbbm{1}_{\left\{\sigma_N (t)=I\right\}}}{N\left|S_N\right|} \right\}.
	\end{equation*}
	
	\noindent The final update step, then, is given by
	
	\begin{equation}
	\mathbf{v}^{\tau+\dt} = \mathbf{v}_{\text{\tiny imp}}^{\tau+\dt} + \mathbf{v}_{\text{\tiny exp}}^{\tau+\dt}.
	\end{equation}

	\pagebreak

	\section*{Proportion of infections from the cell-to-cell mode of spread}
	
	\begin{figure}[h!]
		\centering
		\includegraphics[width=0.6\linewidth]{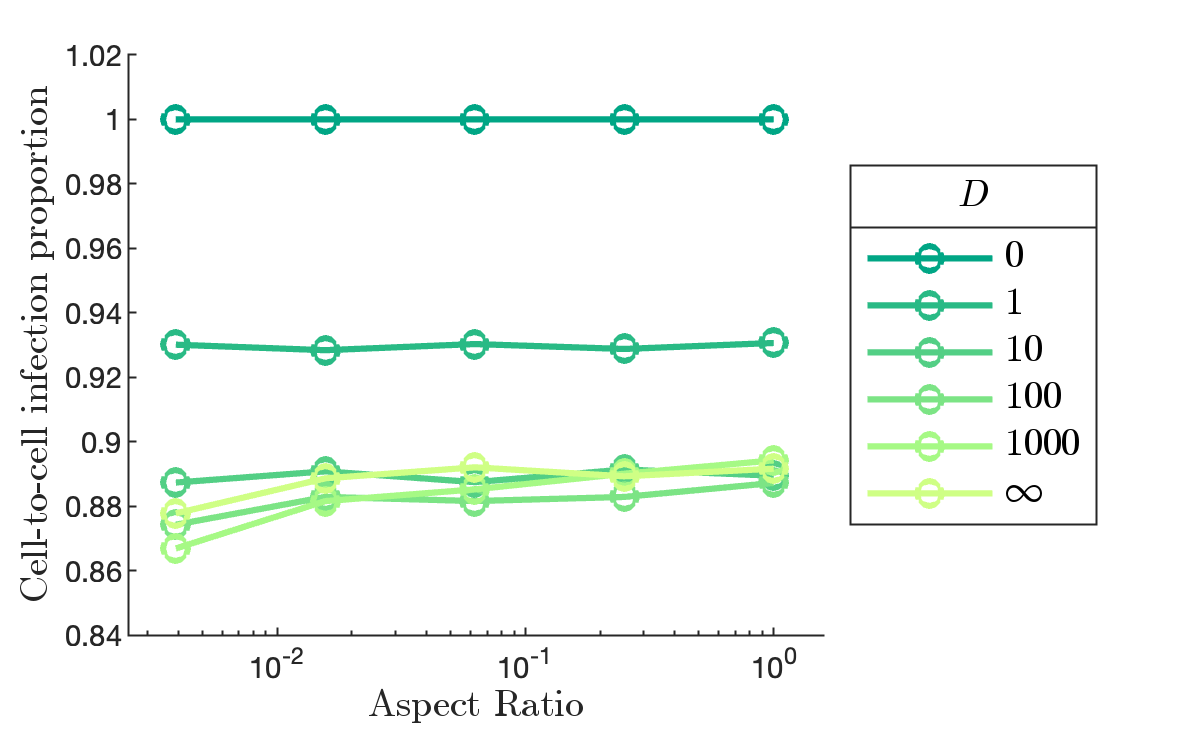}
		\caption{Proportion of infections arising from the cell--to--cell route for infections on tube-shaped tissues of specified aspect ratio and with given viral diffusion coefficient. Accompanies Figure 2 in the main text.}
		\label{fig:Pcc_vs_AR}
	\end{figure}

	\section*{Lineage competition and extinction on a branching tissue --- source on open edge \emph{[video]}}
	
	\begin{video}[h!]
		\centering
		DOI:10.26188/28537028
		\caption{Simulation of infection of a branching tissue by four viral lineages, seeded on the open edge of the tissue. Target cells are shown in light grey, infected cells are coloured by the lineage which infected it, and dead cells are shown in dark grey. Accompanies Figure 6(e) in the main text.}
	\end{video}

	\section*{Lineage competition and extinction on a branching tissue --- source on branching edge \emph{[video]}}
	
	\begin{video}[h!]
		\centering
		DOI:10.26188/28537061
		\caption{Simulation of infection of a branching tissue by four viral lineages, seeded on the branching edge of the tissue. Target cells are shown in light grey, infected cells are coloured by the lineage which infected it, and dead cells are shown in dark grey. Accompanies Figure 6(f) in the main text.}
	\end{video}

	\pagebreak

	\section*{Immune response parameter sweep --- choice of contours}
	
	\begin{figure}[h!]
		\centering
		\includegraphics[width=0.7\linewidth]{all_contours_comparison}
		\caption{$\Finf$ contours on $\tact$--$\Fthresh$ space for infections of the branching tree seeded in Generation 1 or 5. Immune parameters as defined in the main text.}
		\label{fig:all_Finf_contours_comparison}
	\end{figure}

\pagebreak

	\section*{Anatomically accurate respiratory tract dimensions}
	
	\begin{figure}[h!]
		\begin{subfigure}{0.48\textwidth}
			\centering
			\raisebox{7em}{(a)}
			\includegraphics[height=5cm]{bronchi_unwrapped}
			\end{subfigure}
		\begin{subfigure}{0.48\textwidth}
			\centering
						\raisebox{7em}{(b)}
			\includegraphics[height=5cm]{alveoli_unwrapped}
			\end{subfigure}\\[1em]
		\begin{subfigure}{0.48\textwidth}
			\centering
			\renewcommand{\arraystretch}{1.2}
			{\scriptsize \color{black}
			\begin{tabular}{c|c|c|c|c|c}
				Generation&1&2&3&4&5\\
				\hline
				Circ. (mm)&56.6&38.3&26.1&17.6&14.1\\
				\hline
				Length (mm)&120&47.6&19.0&17.6&12.7
			\end{tabular}
		}
		\end{subfigure}
		\begin{subfigure}{0.48\textwidth}
			\centering
			\renewcommand{\arraystretch}{1.2}
			{\scriptsize \color{black}
			\begin{tabular}{c|c|c|c|c|c}
				Generation&1&2&3&4&5\\
				\hline
				Circ. (mm)&1.57&1.48&1.41&1.35&1.29\\
				\hline
				Length (mm)&1.17&0.99&0.83&0.70&0.59
			\end{tabular}
		}
		\end{subfigure}
		\caption{Unwrapped airway tissue dimensions based on experimental measurements of the human respiratory tree \cite{makevnina_lung_atlas}, where we take five consecutive generations from (a) the top of the respiratory tree and (b) the terminal five generations before the alveolar sacs. \addednew{Measurements used to generate the figures are given in tables.}}
		\label{fig:anatomically_accurate_lungs}
	\end{figure}

	\clearpage
\bibliographystyle{apalike}